\documentclass[twocolumn,prb,superscriptaddress]{revtex4-1}
 
\usepackage{graphicx,amsmath,xcolor,soul}
\usepackage[capitalise]{cleveref} 
\usepackage{xcolor} 
\usepackage{amsfonts}

\newcommand{\FG}[1]{{\color{blue}#1}}

\def\oden{Oden Institute for Computational Engineering and Sciences,
The University of Texas at Austin, Austin, Texas 78712, USA}
\def\physics{Department of Physics, The University of Texas at Austin, Austin, Texas 78712, USA}
\def\macau{Institute of Applied Physics and Materials Engineering,
University of Macau, Macao SAR 999078, P. R. China}
\def\ubc{Fisika Saila, University of the Basque Country UPV/EHU, 48080 Bilbao, Basque 
Country, Spain}
\def\dipc{Donostia International Physics Center (DIPC), Paseo Manuel de Lardizabal 4, 20018 Donostia-San 
Sebasti\'{a}n, Spain}
\def\EHUq{EHU Quantum Center, University of the Basque Country UPV/EHU, Barrio Sarriena, s/n, 48940 Leioa, Biscay, Spain}

\begin{document}

\title{\textit{Ab initio} self-consistent many-body theory of polarons at all couplings}

\author{Jon Lafuente-Bartolome}
\author{Chao Lian}
  \affiliation{\oden}
  \affiliation{\physics}
\author{Weng Hong Sio}
  \affiliation{\macau}
\author{Idoia G. Gurtubay}
\author{Asier Eiguren}
  \affiliation{\ubc}
  \affiliation{\dipc}
  \affiliation{\EHUq}
\author{Feliciano Giustino}
  \email{fgiustino@oden.utexas.edu}
  \affiliation{\oden}
  \affiliation{\physics}

\date{\today}

\begin{abstract}
We present a theoretical framework to describe polarons from first principles 
within a many-body Green's function formalism. Starting from a general electron-phonon Hamiltonian,
we derive a self-consistent Dyson equation in which the phonon-mediated self-energy is composed by 
two distinct terms. One term is the Fan-Migdal self-energy and describes dynamic
electron-phonon processes, the other term is a new contribution to the self-energy originating from the 
static displacements of the atomic nuclei in the polaronic ground state. 
The lowest-order approximation to the present theory yields the standard many-body 
perturbation theory approach to electron-phonon interactions in the limit of large
polarons, and the \textit{ab initio} polaron equations introduced in [Sio \textit{et al.},
Phys. Rev.  B 99, 235139 (2019); Phys. Rev. Lett. 122, 246403 (2019)]
in the limit of small polarons. 
A practical recipe to implement the present unifying formalism in first-principles calculations 
is outlined. We apply our method to the Fr{\" o}hlich model, 
and obtain remarkably accurate polaron energies at all couplings, 
in line with Feynman's polaron theory
and diagrammatic Monte Carlo calculations. 
We also recover the well-known results of Fr\"ohlich and Pekar at weak and strong coupling, respectively. 
The present approach enables predictive many-body calculations of polarons in real materials at all couplings.
\end{abstract}

\maketitle

\section{Introduction}

A charge carrier propagating through a crystal may induce distortions in the lattice through 
the electron-phonon interaction.
The quasiparticle formed by the carrier and the lattice distortion is referred to as a polaron.\cite{AlexandrovMott1996,Alexandrov2008,Emin2012,Devreese2020arXiv}  
The ionic displacements surrounding 
the carrier may lead to an increase of its effective mass, and, in the case of strong 
electron-phonon coupling, may ultimately form a potential well in which the polaron becomes 
self-trapped.\cite{LandauPZS1933,PekarZETF1946}

A detailed characterization of polarons in materials has been possible throughout the last 
decades by a combination of an array of experimental techniques.\cite{FranchiniNREVMAT2021}
Polarons have been proposed to play a crucial role in the exotic properties of several quantum 
materials, such as high-temperature cuprate superconductors,\cite{ZhaoNAT1997} colossal 
magnetoresistance manganites,\cite{TeresaNAT1997} and halide perovskites.\cite{MiyataSADV2017}
In particular, the low-energy satellites observed in angle-resolved photoemission 
spectroscopy (ARPES) experiments are considered the hallmark
of polarons in doped oxides.\cite{MoserPRL2013,ChenNCOM2015,WangNMAT2016,CancellieriNCOM2016,VerdiNATCOMM2017,Riley2018}
It is generally accepted that polarons govern the
transport,\cite{ZhangJAP2007} optical,\cite{vanMechelenPRL2008} and chemical properties of 
conducting oxides.\cite{ReticcioliPRL2019}

On the theoretical side, the study of polarons mostly focused on idealized models such as 
the Fr{\" o}hlich \cite{FrohlichPHM1950,FrohlichADP1954, Devreese2020arXiv} and the Holstein 
\cite{HolsteinAP1959_I,HolsteinAP1959_II} models, as well as the Su-Schrieffer-Heeger 
model.\cite{SSHPrl1979} 
These models have provided a fertile playground for the development and 
application of advanced many-body techniques \cite{Alexandrov2010} such as variational path 
integral methods,\cite{FeynmanPR1955,OsakaPTP1959} diagrammatic Monte Carlo,\cite{ProkofevPRL1998,
MischenkoPRB2000} dynamical mean field theory,\cite{CiuchiPRB1997,FratiniPRL2003} 
and renormalization group approaches.\cite{GrusdtPRB2016}
While these methods are of great fundamental interest, they are not directly applicable
to the study of polarons in real materials.

\raggedbottom

Recent developments in density functional theory (DFT), density functional perturbation theory (DFPT), 
and many-body perturbation theory have
opened promising new avenues to study polarons in real materials 
from an \textit{ab initio} perspective.\cite{GiustinoRMP2017,SioPRL2019, SioPRB2019, LeePRM2021}
For instance,
by combining first principles calculations with many-body Green's function techniques,
it has been possible to reproduce the signatures of the electron-phonon interaction in the ARPES spectra of doped semiconductors
to high accuracy.\cite{VerdiNATCOMM2017,Riley2018,GoiricelayaCOMMPHYS2019}
However, at this level of theory,
the possibility of spatial correlations between electrons and phonons is not taken into account,\cite{GiustinoRMP2017}
since it is generally assumed that, upon electron addition or removal, both the electron and the
phonon subsystems maintain the periodicity of the original crystalline lattice.

An alternative, heuristic approach to model the formation of polarons from first principles 
consists of performing direct DFT calculations on supercells of insulators with an added or removed electron,
and relaxing the structure to seek for distorted configurations which are energetically favorable
with respect to the original periodic structure.\cite{FranchiniPRL2009,DeskinsPRB2007,LanyPRB2009,VarleyPRB2012,SetvinPRL2014,HimmetogluPRB2014,KokottNJP2018,ReticcioliPRX2017}.

In a recent work,\cite{SioPRB2019,SioPRL2019} Sio \textit{et al.} have formalized the DFT approach to 
the polaron problem, replacing supercell calculations by a set of coupled equations
whose ingredients are the electron band structures, phonon dispersions, and electron-phonon matrix elements 
obtained from DFPT calculations in the crystal unit cell. This method established the link
between model Hamiltonian and \textit{ab initio} approaches to the polaron problem,
and makes it possible to study polaron formation in materials in a systematic way.

However, the intrinsic limitations of DFT, such as the adiabatic and classical approximations for the nuclei,
are naturally inherited by the method of Ref.~\onlinecite{SioPRB2019}. As a consequence, this approach
does not capture dynamical renormalization effects that give rise, for example, to the phonon satellites 
in ARPES spectra.

In this work, we generalize the theory of Ref.~\onlinecite{SioPRB2019} to a many-body formalism beyond
density functional theory. In particular,
we present a Green's function theory of electron-phonon interactions
that captures spatial correlations between lattice distortions and electrons in the many-body ground state.
We find that, besides the standard Fan-Migdal self-energy,\cite{GiustinoRMP2017}
one must consider an additional self-energy contribution that arises from the nonvanishing expectation
value of the atomic displacements when the electron is ``pinned'' around a lattice site.

After presenting the general formalism, we discuss approximations that can be used to implement
this methodology in existing \textit{ab initio} codes. 
This analysis allows us to establish the links between our general formalism, 
the DFT polaron equations,\cite{SioPRL2019,SioPRB2019}
and the Allen-Heine theory of band structure renormalization.\cite{AllenHeineJPC1976}

As a first proof of concept, we apply our methodology to the Fr{\"o}hlich model, and we
benchmark our proposed approximations with respect to the all-coupling path-integral method by 
Feynman\cite{FeynmanPR1955}
and diagrammatic Monte Carlo calculations.\cite{ProkofevPRL1998,MischenkoPRB2000}
We show that our theory naturally connects the established results at weak and strong coupling limits,
and predicts polaron energies with
remarkably good accuracy throughout the whole range of couplings. 
Furthermore, as a first \textit{ab initio} demonstration of this method,
we describe in detail the computational procedure that we used to calculate 
the full polaronic renormalization of the band gap in LiF.
The main results of this calculation and its implications 
in the theory of the phonon-mediated renormalization of band structures
are discussed in the companion manuscript.\cite{AccompanyingPaper}


The manuscript is organized as follows.  In Sec.~\ref{sec:theory} we develop our general formalism.
In particular, we introduce the electron-phonon Hamiltonian in Sec.~\ref{sec:elph_ham}.
In Sec.~\ref{sec:green} we apply Schwinger's functional derivative technique
to obtain an equation of motion for the electron Green's function
which can be rewritten as a Dyson equation, and identify two separate self-energy contributions.
In Sec.~\ref{sec:vertex} we introduce an expression for the vertex function
and a related approximation to simplify the equations.
In Sec.~\ref{sec:displacements}
we obtain an expression for the expectation value of the atomic displacement operator
in terms of the electron density, which allows us to make the Dyson equation fully self-consistent.
In Sec.~\ref{sec:lehmann}
we introduce the Lehmann representation of the electron Green's function, which we use
to derive a Schr{\"o}dinger-like equation for the Dyson orbitals describing the electronic
part of the polaron quasiparticle.
In Sec.~\ref{sec:selfen_from_Ank} the self-energies are rewritten in terms of the polaron quasiparticle amplitudes,
and in Sec.~\ref{sec:qp_eq} we present the self-consistent many-body polaron equations.
In Sec.~\ref{sec:toten},
an expression for the total energy in the many-body ground state of the coupled electron-phonon
system is derived. 
In Sec.~\ref{sec:abinitio}, we develop  approximations of the many-body equations to make the
formalism useful for practical \textit{ab initio} calculations.
Transparent links with the DFT polaron equations and the standard self-energies for electron-phonon 
coupling are established in Sec.~\ref{sec:relation_to_Sio}.
In Sec.~\ref{sec:pert}, we outline a practical recipe to implement 
the lowest-order approximation to our theory in \textit{ab initio} calculations.
In Sec.~\ref{sec:frohlich}, we apply our methodology to the Fr{\"o}hlich model,
and report benchmarks against
the well-known weak and strong coupling limits, as well as Feynman's path integral solution 
and diagrammatic Monte Carlo results.
In Sec.~\ref{sec:LiF}, we outline the computational setup that we used to calculate
polarons in LiF, as reported in the companion manuscript Ref.~\onlinecite{AccompanyingPaper}.
In Sec.~\ref{sec:loc} we address the issue of translational invariance of the polaronic solutions,
and we explain how we can ``pin'' the polaron at a given lattice site.
In Sec.~\ref{sec:conclusion} we summarize our key findings and we anticipate possible future 
developments.

\section{Self-consistent Green's function approach to polarons} \label{sec:theory}

\subsection{Electron-phonon Hamiltonian} \label{sec:elph_ham}

The starting point of our derivation is the standard Hamiltonian describing a coupled electron-phonon system \cite{GiustinoRMP2017}:
\begin{eqnarray} \label{eq:elph_ham}
	\hat{H} &=& \hat{H}_{\mathrm{e}} + \hat{H}_{\mathrm{p}} + \hat{H}_{\mathrm{ep}} \nonumber \\
        &=& \sum_{n\textbf{k}} \varepsilon_{n\mathbf{k}} \hat{c}_{n\mathbf{k}}^\dagger \hat{c}_{n\mathbf{k}} +
        \sum_{\mathbf{q}\nu} \hbar\omega_{\mathbf{q}\nu} (\hat{a}_{\mathbf{q}\nu}^\dagger \hat{a}_{\mathbf{q}\nu}+1/2) \nonumber \\
        &+& N_p^{-\frac{1}{2}} \sum_{\substack{\mathbf{k},\mathbf{q} \\ m n \nu }} g_{mn\nu}(\mathbf{k},\mathbf{q}) \,
        \hat{c}_{m\mathbf{k}+\mathbf{q}}^\dagger \hat{c}_{n\mathbf{k}} (\hat{a}_{\mathbf{q}\nu}+\hat{a}_{-\mathbf{q}\nu}^\dagger) ~,\hspace{0.5cm}
\end{eqnarray}
where $\varepsilon_{n\mathbf{k}}$ is the single-particle eigenvalue of an electron in the band $n$ with crystal momentum $\mathbf{k}$,
$\omega_{\mathbf{q}\nu}$ is the frequency of a phonon in the branch $\nu$ with crystal momentum $\mathbf{q}$,
and $\hat{c}^\dagger_{n\mathbf{k}}/\hat{c}_{n\mathbf{k}}$
($\hat{a}^\dagger_{\mathbf{q}\nu}/\hat{a}_{\mathbf{q}\nu}$) are the associated fermionic (bosonic) creation/annihilation operators.
The electron-phonon coupling matrix elements are represented by $g_{mn\nu}(\mathbf{k},\mathbf{q})$,
and $N_p$~is the number of unit cells in the periodic Born-von K\'arm\'an (BvK) supercell.
To make the following derivations more compact, we introduce the complex normal coordinate 
operator,\cite{GiustinoRMP2017}
\begin{equation} \label{eq:zqnu}
    \hat{z}_{\mathbf{q}\nu} = \nolinebreak  \sqrt{\frac{\hbar}{2M_{0}\omega_{\mathbf{q},\nu}}} \, ( \hat{a}_{\mathbf{q}\nu}+\nolinebreak \hat{a}_{-\mathbf{q}\nu}^\dagger )~,
\end{equation}
where $M_0$ is a reference mass. The operator $\hat{z}_{\mathbf{q}\nu}$ has dimensions of a length.
We note that Eq.~(\ref{eq:elph_ham}) is an effective Hamiltonian,
where
we assume that electron-electron interaction effects have been incorporated in the single-particle energies $\varepsilon_{n\mathbf{k}}$,
so that electrons can be identified as well-defined quasiparticles in the absence of electron-phonon coupling.
Moreover,
phonons are described in the harmonic approximation,
only linear electron-phonon coupling is retained,
and the phonon frequencies and the electron-phonon matrix elements
already incorporate electronic screening at a mean-field level.
In practical \textit{ab initio} calculations,
the electron energies are typically obtained via DFT or GW calculations,\cite{HybertsenLouiePRB1986}
and phonon frequencies and electron-phonon matrix elements are obtained from DFPT calculations.\cite{BaroniRMP2001}
The use of Eq.~(\ref{eq:elph_ham}) to compute most of the physical observables related to the
renormalization of electrons due to the
electron-phonon interaction,
such as for instance temperature-dependent band structures,
can be justified rigorously by starting from a more general electron-ion Hamiltonian.\cite{GiustinoRMP2017}
The study of phonon renormalization requires more care\cite{GiustinoRMP2017},
and it is not attempted in this work.
Most model Hamiltonian approaches to the polaron problem,
such as the Fr{\"o}hlich \cite{FrohlichADP1954} (see Sec.~\ref{sec:frohlich}) or the Holstein \cite{HolsteinAP1959_I,HolsteinAP1959_II} model,
are based on further simplifications of Eq.~(\ref{eq:elph_ham}).

As we discuss in detail in Sec.~\ref{sec:loc},
an additional term is needed in Eq.~\eqref{eq:elph_ham} to break translational symmetry and 
pin the polaron at a given lattice site. 
In the following we omit this term for clarity,
since it does not alter the final results, 
and we return to it in Sec.~\ref{sec:loc}.

\subsection{Equation of motion for the electron Green's function} \label{sec:green}

The central object in our derivation is the electron Green's function,
which is defined as:
\begin{equation} \label{eq:Green_def_r}
    G(\mathbf{r}t,\mathbf{r}'t') = -\frac{i}{\hbar} \langle N+1 | \, \hat{T} \, 
                                    \hat{\psi}(\mathbf{r},t) \, \hat{\psi}^{\dagger}(\mathbf{r}',t')
                                    \, | N+1 \rangle\FG{,}
\end{equation}
where $|N+1\rangle$ represents the many-body ground state of an $(N+1)$-electron system, 
$\hat{T}$ is the time-ordering operator,\cite{FetterWalecka2003}
and $\hat{\psi}^{\dagger}/\hat{\psi}$ are the electron field creation/annihilation operators.
In the following we consider the electron polaron for definiteness, but our results hold
unchanged for hole polarons. 
The field operators can be written in the single-particle basis used in Eq.~(\ref{eq:elph_ham}),
\begin{eqnarray}
\label{eq:fieldops_singleparticle}
        \hat{\psi}(\mathbf{r}) &=& \sum_{n\mathbf{k}} \psi_{n\mathbf{k}}(\mathbf{r}) \, \hat{c}_{n\mathbf{k}} , 
\end{eqnarray}
being $\psi_{n\mathbf{k}}(\mathbf{r})$ the single-particle Bloch wave functions,
so that the Green's function in the single-particle basis reads:
\begin{equation} \label{eq:Green_def}
    G_{n\mathbf{k},n'\mathbf{k'}} (t,t') = -\frac{i}{\hbar} \langle N+1 | \, \hat{T} \, \hat{c}_{n\mathbf{k}}(t) \, \hat{c}^\dagger_{n'\mathbf{k'}}(t') \, | N+1 \rangle ~.
\end{equation}
In Eqs.~(\ref{eq:Green_def_r}) and (\ref{eq:Green_def}), $|N+1\rangle$ represents the polaronic many-body ground state,
which corresponds to a single electron added to semiconductor or insulator with filled valence bands and
empty conduction bands, correlated with its accompanying phonon cloud.
In the following, the brackets $\langle \, \rangle$ represent
the expectation value of operators over the $|N+1\rangle$ state,
unless otherwise specified.
An important distinction from previous Green's function approaches to the polaron problem \cite{ProkofevPRL1998,MischenkoPRB2000}
is that in those studies the expectation value in the definition of Eq.~(\ref{eq:Green_def}) is taken over the 
ground state of the $N$-electron system, i.e. the system in absence of the extra electron.
Our present choice of starting from the $|N+1\rangle$ state is useful to better connect with DFT calculations,
as it will become clear shortly. We elaborate further on this point in Sec.~\ref{sec:conclusion}.

The time dependence in the electron operators can be described within the Heisenberg picture,
so that their equation of motion is given by:
\begin{eqnarray} \label{eq:eqmo_cnk}
    &i\hbar& \frac{\partial}{\partial t} \hat{c}_{n\mathbf{k}}(t) = \left[ \hat{c}_{n\mathbf{k}}(t), \hat{H} \right] 
    = \varepsilon_{n\mathbf{k}} \hat{c}_{n\mathbf{k}}(t) \nonumber \\
    &+& \sqrt{\frac{2\,M_{0}\,\omega_{\mathbf{q},\nu}}{\hbar N_p}} \! \sum_{n' \mathbf{q} \nu } \! g_{nn'\nu}(\mathbf{k-q},\mathbf{q}) \,
    \hat{c}_{n'\mathbf{k-q}}(t) \, \hat{z}_{\mathbf{q}\nu}(t) ,
\end{eqnarray}
where the anticommutation relations for the electron operators,
$\{ \hat{c}_{n\mathbf{k}}, \hat{c}^{\dagger}_{n'\mathbf{k'}} \} = \delta_{n\mathbf{k},n'\mathbf{k'}}$ and
$\{ \hat{c}_{n\mathbf{k}}, \hat{c}_{n'\mathbf{k'}} \} = \{ \hat{c}^{\dagger}_{n\mathbf{k}}, \hat{c}^{\dagger}_{n'\mathbf{k'}} \} = 0$
have been used.
Combining Eqs.~(\ref{eq:Green_def}) and (\ref{eq:eqmo_cnk}),
the following equation of motion for the electron Green's function is obtained:
\begin{align} \label{eq:eqmo_green}
        & \left ( i\hbar \frac{\partial}{\partial t} - \varepsilon_{n\mathbf{k}} \right ) G_{n\mathbf{k},n'\mathbf{k'}} (t,t') = 
        \delta(t-t') \, \delta_{n\mathbf{k},n'\mathbf{k'}} \nonumber \\
        &\hspace{25pt} - \frac{i}{\hbar} N_p^{-\frac{1}{2}} \sum_{n'' \mathbf{q} \nu } g_{nn''\nu}(\mathbf{k-q},\mathbf{q}) \nonumber \\
        &\hspace{25pt} \times \sqrt{\frac{2\,M_{0}\,\omega_{\mathbf{q},\nu}}{\hbar}}
        \langle \, \hat{T} \, \hat{z}_{\mathbf{q}\nu}(t) \, \hat{c}_{n''\mathbf{k-q}}(t) \, \hat{c}^\dagger_{n'\mathbf{k'}}(t') \, \rangle ~.
\end{align}
In order to deal with the last term of Eq.~(\ref{eq:eqmo_green}),
we proceed with Schwinger's functional derivative technique.\cite{Schwinger1951,KatoPTP1960}
The main idea is to add an external source term that couples to the normal mode coordinates via:
\begin{equation} \label{eq:source_term}
    \hat{H}_{\mathrm{ext}}(t) = \sum_{\mathbf{q\nu}} F_{\mathbf{q}\nu}(t) \, \hat{z}_{\mathbf{q}\nu}(t) ~.
\end{equation}
This term will be set to zero at the end of the derivation, but it is instrumental to
obtain a set of self-consistent equations for the electron Green's function
by taking functional derivatives with respect to the fictitious forces $F_{\mathbf{q}\nu}(t)$.
Furthermore, 
this term is needed to break translational symmetry and pin the polaron around a lattice site
(see Sec.~\ref{sec:loc}).
The Schwinger's functional derivative technique has proven very successful in electronic structure
theory, and is at the heart of all modern developments in the GW method.\cite{OnidaRMP2002,ReiningWIREs2018,GolzeFC2019}

We rewrite Eq.~(\ref{eq:eqmo_green}) using the following functional identity, first derived in 
Ref.~\onlinecite{KatoPTP1960} and employed extensively in 
Refs.~\onlinecite{HedinLundqvistSSP1969,EngelsbergSchriefferPR1963}:
\begin{widetext}
\begin{equation} \label{eq:Kato_identity}
    \frac{\delta \langle \, \hat{T} \, \hat{O}_{1}(t_{1}) \hat{O}_{2}(t_{2}) \dots \, \rangle}{\delta F_{\mathbf{q}\nu}(t)}
    = -\frac{i}{\hbar} \langle \, \hat{T} \, \hat{z}_{\mathbf{q}\nu}(t) \, \hat{O}_{1}(t_{1}) \hat{O}_{2}(t_{2}) \dots \, \rangle
    +\frac{i}{\hbar} \langle \hat{z}_{\mathbf{q}\nu}(t) \rangle \langle \, \hat{T} \, \hat{O}_{1}(t_{1}) \hat{O}_{2}(t_{2}) \dots \, \rangle ~.
\end{equation}
Here, $\hat O$ represents a generic many-body operator. Using this expression, Eq.~(\ref{eq:eqmo_green}) becomes:
\begin{align} \label{eq:eqmo_green_2}
    \left ( i\hbar \frac{\partial}{\partial t} - \varepsilon_{n\mathbf{k}} \right ) G_{n\mathbf{k},n'\mathbf{k'}} (t,t') 
    = &~ \delta(t-t') \, \delta_{n\mathbf{k},n'\mathbf{k'}} \nonumber \\
    &+ N_p^{-\frac{1}{2}}\!\! \sum_{n'' \mathbf{q} \nu } g_{nn''\nu}(\mathbf{k-q},\mathbf{q}) \sqrt{\frac{2\,M_{0}\,\omega_{\mathbf{q},\nu}}{\hbar}}
    \left ( i\hbar \frac{\delta}{\delta F_{\mathbf{q}\nu}(t)} + \langle \,  \hat{z}_{\mathbf{q}\nu}(t) \, \rangle \right )
    G_{n''\mathbf{k-q},n'\mathbf{k'}} (t,t') ~.
\end{align}
To eliminate the dependence on $F_{\mathbf{q}\nu}(t)$,
we first rewrite the functional derivative in the last term
in terms of the inverse of the Green's function,\cite{KadanoffBaym1962}
\begin{equation} \label{eq:inverse_Green}
    \frac{\delta G_{n\mathbf{k},n'\mathbf{k'}} (t,t')}{\delta F_{\mathbf{q}\nu}(t)}
    = - \int dt'' dt''' \sum_{\substack{ n''\mathbf{k}'' \\ n'''\mathbf{k}''' }}
    G_{n\mathbf{k},n''\mathbf{k}''} (t,t'')
    \, \frac{\delta G^{-1}_{n''\mathbf{k}'',n'''\mathbf{k}'''} (t'',t''')}{\delta F_{\mathbf{q}\nu}(t)}
    \, G_{n'''\mathbf{k}''',n'\mathbf{k}'} (t''',t') ~,
\end{equation}
and then we apply the following functional chain rule,\cite{KadanoffBaym1962}
\begin{equation} \label{eq:chainrule_G_z_phi}
    \frac{\delta G^{-1}_{n\mathbf{k},n'\mathbf{k}'} (t,t')}{\delta F_{\mathbf{q}\nu}(t'')}
    = \! \int \! dt'''  \sum_{\mathbf{q}' \nu'} 
    \frac{\delta G^{-1}_{n\mathbf{k},n'\mathbf{k}'} (t,t')}{\delta \langle \,  \hat{z}_{\mathbf{q}' \nu'}(t''') \, \rangle}
    \frac{\delta \langle \,  \hat{z}_{\mathbf{q}' \nu'}(t''') \, \rangle }{\delta F_{\mathbf{q}\nu}(t'')} ~.
\end{equation}
The last term on the right hand side can be identified with the phonon Green's function, which is given by:
\begin{equation} \label{eq:phonon_Green1}
    D_{\mathbf{q}'\nu',\mathbf{q}\nu}(t',t) =
    -i\sqrt{\frac{2M_{0}\omega_{\mathbf{q}\nu}}{\hbar}} \sqrt{\frac{2M_{0}\omega_{\mathbf{q}'\nu'}}{\hbar}}
     \langle \, \hat{T} \left[ \hat{z}_{\mathbf{q}'\nu'}(t') - \langle \hat{z}_{\mathbf{q}'\nu'}(t') \rangle  \right]
   \left[ \hat{z}_{\mathbf{q}\nu}(t) - \langle \hat{z}_{\mathbf{q}\nu}(t) \rangle  \right] 
   \, \rangle~.
\end{equation}
In fact, by using this definition inside Eq.~(\ref{eq:Kato_identity}), we find:
\begin{equation} \label{eq:phonon_Green2}
    \frac{\delta \langle \,  \hat{z}_{\mathbf{q}\nu}(t) \, \rangle}{\delta F_{\mathbf{q'}\nu'}(t')}
    =  \frac{1}{\sqrt{2M_0 \omega_{\mathbf{q}\nu}}} \frac{1}{\sqrt{2M_0 \omega_{\mathbf{q}'\nu'}}}
    D_{\mathbf{q}\nu,\mathbf{q}'\nu'}(t,t')~.
\end{equation}
Now we define the vertex function as:
\begin{equation} \label{eq:vertex_definition}
    \Gamma_{n\mathbf{k},n'\mathbf{k}',\nu\mathbf{q}}(t,t',t'')
    = - \sqrt{\frac{\hbar}{2\,M_{0}\,\omega_{\mathbf{q} \nu}}}
    \, \frac{\delta G^{-1}_{n\mathbf{k},n'\mathbf{k}'} (t,t')}{\delta \langle \,  \hat{z}_{\mathbf{q}\nu}(t'') \, \rangle} ~,
\end{equation}
and we write the non-interacting Green's function as:
\begin{equation} \label{eq:nonint_G0}
    (G^{0})^{-1}_{n\mathbf{k},n'\mathbf{k'}}(t,t')
    = \left ( i\hbar \frac{\partial}{\partial t} - \varepsilon_{n\mathbf{k}} \right ) 
    \, \delta(t-t') \, \delta_{n\mathbf{k},n'\mathbf{k'}} ~,
\end{equation}
Using the last two relations together with Eqs.~(\ref{eq:inverse_Green})-(\ref{eq:phonon_Green2}),
we can rewrite Eq.~(\ref{eq:eqmo_green_2}) as a Dyson equation:
    \begin{equation} \label{eq:dyson_t}
        \int \! dt'' \sum_{n''\mathbf{k''}} 
        \left[ (G^{0})^{-1}_{n\mathbf{k},n''\mathbf{k''}}(t,t'') - 
        \Sigma^{\mathrm{P}}_{n\mathbf{k},n''\mathbf{k''}}(t,t'') - \Sigma^{\mathrm{FM}}_{n\mathbf{k},n''\mathbf{k''}}(t,t'') \right]
        G_{n''\mathbf{k''},n'\mathbf{k'}}(t'',t')
        = \delta(t-t') \, \delta_{n\mathbf{k},n'\mathbf{k'}}~,
    \end{equation}
where the self-energies $\Sigma^{\mathrm{P}}$ and $\Sigma^{\mathrm{FM}}$ 
are defined as follows:
\begin{eqnarray}
    \Sigma^{\mathrm{FM}}_{n\mathbf{k},n'\mathbf{k'}}(t,t')
    &=&~
    i N_p^{-1/2}
    \int \! dt'' dt''' \! \sum_{n'' \mathbf{k''}}
    \sum_{n'''\mathbf{k}'''}
    \sum_{\nu \nu' \mathbf{q}'}
    g_{n n'' \nu}(\mathbf{k}'',\mathbf{k-k''}) 
    G_{n''\mathbf{k}'',n'''\mathbf{k'''}}(t,t''')
    \nonumber \\
    && \hspace{40pt}
    \times
    \,\Gamma_{n'''\mathbf{k'''},n'\mathbf{k'},\mathbf{q}'\nu'}(t''',t',t'')
    \,D_{\mathbf{q}'\nu',\mathbf{k-k''}\nu}(t'',t)
    \label{eq:selfen_FM}~,  \\[10pt]
    \Sigma^{\mathrm{P}}_{n\mathbf{k},n'\mathbf{k'}}(t,t')
    &=&~ N_p^{-1/2}
    \sum_{\nu } \sqrt{\frac{2\,M_{0}\,\omega_{\mathbf{k-k'},\nu}}{\hbar}}\, g_{nn'\nu}(\mathbf{k'},\mathbf{k-k'})
    \, \langle \,  \hat{z}_{\mathbf{k-k'}\nu}(t) \, \rangle \delta(t-t') ~, \label{eq:selfen_LP}
\end{eqnarray}   
\end{widetext}
and the superscripts ``FM'' and ``P'' stand for Fan-Migdal and Polaronic, respectively.
Note that $(G^{0})^{-1}_{n\mathbf{k},n'\mathbf{k'}}(t,t')$,
and $\Sigma_{n\mathbf{k},n'\mathbf{k'}}(t,t')$ have units of energy divided by time.

\begin{figure*}[t]
        \includegraphics[width=1.0\linewidth]{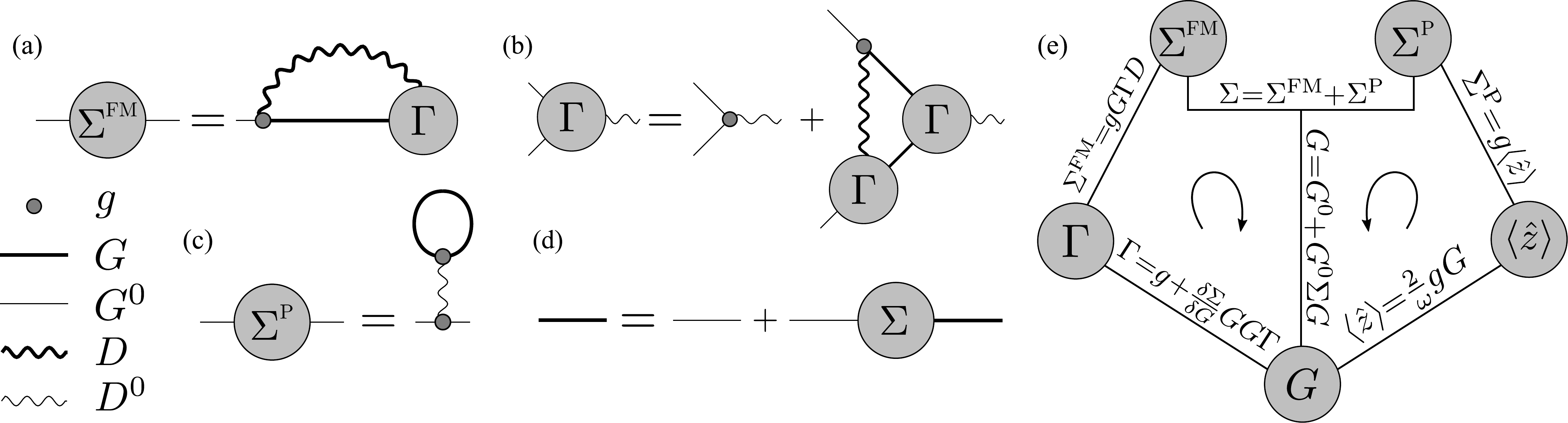}
        \caption{Diagrammatic representation of the self-consistent Green's function theory of polarons.
        Legends for the different parts of the diagrams are given in the lower left corner.
        (a) Fan-Migdal self-energy in Eq.~(\ref{eq:selfen_FM}).
        (b) Self-consistent definition of the vertex function in Eq.~(\ref{eq:vertex_selfcon}).
        (c) Polaronic self-energy in Eq.~(\ref{eq:LP_from_G}).
        (d) Dyson equation, Eq.~(\ref{eq:dyson_t}).
        (e) Schematic representation of the self-consistent solution 
        of Eqs.~(\ref{eq:dyson_t}), (\ref{eq:selfen_FM}), (\ref{eq:selfen_LP}), (\ref{eq:vertex_selfcon}) and (\ref{eq:part_sol_ep}).
        \label{fig:diagrams}}
    \end{figure*}

A diagrammatic representation of the self-energy $\Sigma^{\rm FM}$ in Eq.~(\ref{eq:selfen_FM}) is given in Fig.~\ref{fig:diagrams}(a).
In Sec.~\ref{sec:approx} we show how, using a standard approximation for the vertex $\Gamma$,
we can identify $\Sigma^{\rm FM}$ in Eq.~(\ref{eq:selfen_FM}) with the standard
Fan-Migdal (FM) self-energy \cite{GiustinoRMP2017}.
This self-energy has been key to interpret the spectral kinks and satellites 
observed in photoemission experiments,\cite{MoserPRL2013,VerdiNATCOMM2017,KangNATMAT2018, EigurenPRB2009,GoiricelayaCOMMPHYS2019}
and leads to the
Allen-Heine\cite{AllenHeineJPC1976} theory of band structure renormalization 
including its non-adiabatic generalizations
\cite{AllenPRB1978,MariniPRL2008,GiustinoPRL2010,GonzeAP2011,PonceJCP2015,NeryPRB2016,CarusoPRB2019,BrownPRB2020,MiglioNPJ2020}
upon performing 
the pertinent approximations (see e.g. Ref.~\onlinecite{GiustinoRMP2017}).

To the best of our knowledge, the self-energy
$\Sigma^{\rm P}$ in Eq.~(\ref{eq:selfen_LP})
appeared in Refs.~\onlinecite{EngelsbergSchriefferPR1963} and \onlinecite{MariniPRB2015}, but the connection to polaron formation was not appreciated.
This new self-energy
accounts for the static renormalization of electron energies due to localization effects.
In fact, as we show below, this self-energy reproduces
the Pekar solution to the Fr{\"o}hlich polaron problem in the limit of strong coupling.\cite{PekarZETF1946}

    %

\subsection{Vertex function} \label{sec:vertex}

In order to obtain a closed self-consistent set of equations,
we need to express the vertex function $\Gamma$ in Eq.~(\ref{eq:vertex_definition})
in terms of the electron Green's function. To this aim, we first
invert the Dyson equation in Eq.~(\ref{eq:dyson_t}),
\begin{equation} \label{eq:Dyson_inverted}
    G^{-1}_{n\mathbf{k},n'\mathbf{k'}}(t,t')
    =
    (G^{0})^{-1}_{n\mathbf{k},n'\mathbf{k'}}(t,t') - \Sigma_{n\mathbf{k},n'\mathbf{k'}}(t,t') ~,
\end{equation}
so that we can take the functional derivatives for each term separately.
In the remainder of this section we use numbered indices for convenience.

From Eq.~(\ref{eq:nonint_G0}), we see that $\delta (G^{0})^{-1} / \delta \langle \hat{z} \rangle = 0$.
The functional derivatives of $\Sigma^{\rm P}$ and $\Sigma^{\rm FM}$ are obtained as follows.
For the polaronic self-energy in Eq.~(\ref{eq:selfen_LP}), we find:
\begin{multline} \label{eq:SigmaLP_deriv}
    \frac{\delta \Sigma^{\mathrm{P}}_{n_{1}\mathbf{k}_{1},n_{2}\mathbf{k}_{2}} (t_{1},t_{2})}{\delta \langle \,  \hat{z}_{\mathbf{q}_{3}\nu_{3}}(t_{3}) \, \rangle}
    =
    \delta(t_{1}-t_{2}) \delta(t_{1}-t_{3}) \delta_{\mathbf{q}_{3},\mathbf{k}_{1} - \mathbf{k}_{2}} \\
    \times N_p^{-\frac{1}{2}}
    \sqrt{\frac{2\,M_{0}\,\omega_{\mathbf{q}_{3},\nu}}{\hbar}}
    g_{n_{1}n_{2}\nu_{3}}(\mathbf{k}_{2},\mathbf{q}_{3}) ~.
\end{multline}
For the FM self-energy, we apply the chain rule as in Eq.~(\ref{eq:chainrule_G_z_phi}),
\begin{multline} \label{eq:chainrule_Sigma_G_z}
    \frac{\delta \Sigma^{\mathrm{FM}}_{n_{1}\mathbf{k}_{1},n_{2}\mathbf{k}_{2}} (t_{1},t_{2})}{\delta \langle \,  \hat{z}_{\mathbf{q}_{3}\nu_{3}}(t_{3}) \, \rangle}
    = \int \! dt_{4} dt_{5}  \! \sum_{\substack{n_{4} \mathbf{k}_{4} \\ n_{5} \mathbf{k}_{5}}} \! 
    \frac{\delta \Sigma^{\mathrm{FM}}_{n_{1}\mathbf{k}_{1},n_{2}\mathbf{k}_{2}} (t_{1},t_{2})}{\delta G_{n_{4}\mathbf{k}_{4},n_{5}\mathbf{k}_{5}} (t_{4},t_{5})} \\
    \times \frac{\delta G_{n_{4}\mathbf{k}_{4},n_{5}\mathbf{k}_{5}} (t_{4},t_{5})}{\delta \langle \,  \hat{z}_{\mathbf{q}_{3}\nu_{3}}(t_{3}) \, \rangle}~.
\end{multline}
This expression can be simplified using $G^{-1}$ as in Eq.~\eqref{eq:inverse_Green}:
\begin{align} \label{eq:G_z_inverse}
    \frac{\delta G_{n_{4}\mathbf{k}_{4},n_{5}\mathbf{k}_{5}} (t_{4},t_{5})}{\delta \langle \,  \hat{z}_{\mathbf{q}_{3}\nu_{3}}(t_{3}) \, \rangle}
    = - \int dt_{6} dt_{7} \sum_{\substack{ n_{6}\mathbf{k}_{6} \\ n_{7}\mathbf{k}_{7} }}
    G_{n_{4}\mathbf{k}_{4},n_{6}\mathbf{k}_{6}} (t_{4},t_{6}) \nonumber \\
    \times \frac{\delta G^{-1}_{n_{6}\mathbf{k}_{6},n_{7}\mathbf{k}_{7}} (t_{6},t_{7})}{\delta \langle \,  \hat{z}_{\mathbf{q}_{3}\nu_{3}}(t_{3}) \, \rangle}
    G_{n_{7}\mathbf{k}_{7},n_{5}\mathbf{k}_{5}} (t_{7},t_{5}) ~.
\end{align}
In this expression we recognize the vertex function appearing in the integrand, in the form of 
Eq.~(\ref{eq:vertex_definition}).
Combining Eqs.~(\ref{eq:vertex_definition}) and (\ref{eq:Dyson_inverted})\FG{-}(\ref{eq:G_z_inverse}),
we arrive at the following self-consistent expression for the vertex function:
\begin{widetext}
\begin{align} \label{eq:vertex_selfcon}
    &\Gamma_{n_{1}\mathbf{k}_{1},n_{2}\mathbf{k}_{2},\nu_{3}\mathbf{q}_{3}}(t_{1},t_{2},t_{3})
    = 
    \delta(t_{1}-t_{2}) \, \delta(t_{1}-t_{3}) \, \delta_{\mathbf{q}_{3},\mathbf{k}_{1} - \mathbf{k}_{2}} \,
    N_p^{-\frac{1}{2}}
    \, g_{n_{1}n_{2}\nu_{3}}(\mathbf{k}_{2},\mathbf{q}_{3}) \nonumber \\
    &+
    \int dt_{4} dt_{5} dt_{6} dt_{7}
    \sum_{\substack{n_{4} \mathbf{k}_{4} \\ n_{5} \mathbf{k}_{5} \\ n_{6} \mathbf{k}_{6} \\ n_{7} \mathbf{k}_{7} }}
    \frac{\delta \Sigma^{\mathrm{FM}}_{n_{1}\mathbf{k}_{1},n_{2}\mathbf{k}_{2}} (t_{1},t_{2})}{\delta G_{n_{4}\mathbf{k}_{4},n_{5}\mathbf{k}_{5}} (t_{4},t_{5})}
    G_{n_{4}\mathbf{k}_{4},n_{6}\mathbf{k}_{6}} (t_{4},t_{6}) 
    G_{n_{7}\mathbf{k}_{7},n_{5}\mathbf{k}_{5}} (t_{7},t_{5})
    \Gamma_{n_{6}\mathbf{k}_{6},n_{7}\mathbf{k}_{7},\nu_{3}\mathbf{q}_{3}}(t_{6},t_{7},t_{3}) ~.
\end{align}
\end{widetext}
A diagrammatic representation of Eq.~(\ref{eq:vertex_selfcon}) is given in Fig.~\ref{fig:diagrams}(b).

\raggedbottom

\subsection{Atomic displacements and polaronic self-energy} \label{sec:displacements}

The expectation value of the normal mode operator
that appears in the polaronic self-energy $\Sigma^{\rm P}$
can be expressed as:\cite{GiustinoRMP2017}
\begin{equation} \label{eq:normal_mode_coord}
    \langle \hat{z}_{\mathbf{q}\nu} \rangle = N_p^{-\frac{1}{2}} \sum_{\kappa \alpha p} e^{-i\mathbf{q}\cdot\mathbf{R}_{p}} \sqrt{\frac{M_{\kappa}}{M_{0}}}
                                                                         e^{*}_{\kappa\alpha,\nu}(\mathbf{q}) 
                                                                         \langle \Delta \hat{\tau}_{\kappa\alpha p} \rangle ~,
\end{equation}
where $\Delta \hat{\tau}_{\kappa\alpha p}$ represents the operator for 
the displacement of the nucleus $\kappa$ in the unit cell $p$ along the cartesian direction $\alpha$,
$e_{\kappa\alpha,\nu}(\mathbf{q})$ is the polarization vector of the phonon branch $\nu$ at momentum $\mathbf{q}$,
$M_{\kappa}$ is the mass of the nucleus $\kappa$,
$M_{0}$ is a reference mass (e.g. the proton mass),
and $\mathbf{R}_{p}$ is the lattice vector of the unit cell $p$.
Equations~(\ref{eq:dyson_t}), (\ref{eq:selfen_LP}) and (\ref{eq:normal_mode_coord}) 
show that, if in the polaron ground state the atoms are displaced from their equilibrium
sites, then there is an additional self-energy term to be added to the standard FM self-energy 
contribution.
In the following,
we show that the value of the atomic displacements
is determined by the ground state electron density,
which in turn can be written self-consistently in terms of the renormalized electron Green's function $G$.

In order to obtain an explicit expression for $\langle \Delta \hat{\tau}_{\kappa\alpha p} \rangle$,
it is convenient to rewrite the electron-phonon interaction term of the Hamiltonian in Eq.~(\ref{eq:elph_ham}) 
as:\cite{GiustinoRMP2017}
\begin{equation} \label{eq:Hep_ham_field}
	\hat{H}_{\mathrm{ep}} = \int d\mathbf{r} \sum_{\kappa\alpha p} \frac{\partial V^{0}_{\mathrm{tot}}(\mathbf{r})}{\partial \tau_{\kappa\alpha p}} \,
	\hat{n}_{\mathrm{e}}(\mathbf{r}) \, \Delta\hat{\tau}_{\kappa\alpha p} ~,
\end{equation}
where $V^{0}_{\mathrm{tot}}(\mathbf{r})$ is the total (electronic plus ionic) electrostatic potential 
in the absence of the excess electron,
and $\hat{n}_{\mathrm{e}}(\mathbf{r})=\hat{\psi}^{\dagger}(\mathbf{r})\hat{\psi}(\mathbf{r})$
is the electron density operator.
Similarly,
the phonon term of the Hamiltonian in Eq.~(\ref{eq:elph_ham}) can be rewritten as:\cite{GiustinoRMP2017}
\begin{equation} \label{eq:Hp_Hf}
	\hat{H}_{\mathrm{p}} \! = \!
    - \! \sum_{\kappa\alpha p} \frac{\hbar^2}{2M_\kappa}\frac{\partial^2}{\partial \tau_{\kappa\alpha p}^2}	
	+ \frac{1}{2} \!\! \sum_{\substack{\kappa\alpha p\\\kappa'\alpha' p'}} C_{\kappa\alpha p,\kappa'\alpha' p'} \Delta \hat{\tau}_{\kappa\alpha p} \Delta \hat{\tau}_{\kappa'\alpha' p'} ~,
\end{equation}
where $C_{\kappa\alpha p,\kappa'\alpha' p'}$ is the matrix of the interatomic force constants.

From Eqs.~(\ref{eq:Hep_ham_field}) and (\ref{eq:Hp_Hf}),
an equation of motion for the displacement operator resembling Newton's equation can be obtained:
\begin{equation} \label{eq:eqmo_disp}
	\frac{d^2}{dt^2} \Delta\hat{\tau}_{\kappa \alpha p}(t)
	= -\frac{1}{\hbar^2} [[\Delta\hat{\tau}_{\kappa \alpha p}(t),\hat{H}],\hat{H}] ~.
\end{equation}
Using the commutation relations
$[\Delta\hat{\tau}_{\kappa\alpha p},\Delta\hat{\tau}_{\kappa'\alpha' p'}]
=[\hat{p}_{\kappa\alpha p},\hat{p}_{\kappa'\alpha' p'}]=0$ 
and $[\Delta\hat{\tau}_{\kappa\alpha p},\hat{p}_{\kappa'\alpha' p'}]=i\hbar\,\delta_{\kappa\alpha p, \kappa'\alpha' p'}$,
where $\hat{p}_{\kappa\alpha p} = -i\hbar \, \partial / \partial \tau_{\kappa\alpha p}$,
and taking the expectation value on the polaron ground state,
we are left with a second-order nonhomogeneous differential equation for $\langle \Delta \hat{\tau}_{\kappa\alpha p} \rangle$:
\begin{align} \label{eq:eqmo_disp_expec}
	\frac{d^2}{dt^2} \langle \Delta\hat{\tau}_{\kappa \alpha p}(t) \rangle
	= &- \sum_{\kappa'\alpha' p'} \frac{C_{\kappa\alpha p,\kappa'\alpha' p'}}{M_\kappa} \langle \Delta \hat{\tau}_{\kappa'\alpha' p'}(t) \rangle \nonumber \\
	&- \frac{1}{M_{\kappa}} \int d\mathbf{r} \frac{\partial V_{\mathrm{tot}}^{0}(\mathbf{r})}{\partial \tau_{\kappa\alpha p}} \, 
    n_{\mathrm{e}}(\mathbf{r})~,
\end{align}
where $n_{\mathrm{e}}(\mathbf{r})=\langle \hat{n}_{\mathrm{e}}(\mathbf{r},t) \rangle$
is the expectation value of the density operator on the ground state,
which is stationary.
The first line of Eq.~(\ref{eq:eqmo_disp_expec}) serves as the complementary homogeneous equation,
whose solution is given by a linear combination of normal vibrational modes,
\begin{equation} \label{eq:disp_homo_final}
	\langle  \Delta\hat{\tau}_{\kappa \alpha p}^{\mathrm{hom}}(t) \rangle
	=
	\sum_{\mathbf{q}\nu} u_{\mathbf{q}\nu}
	\sqrt{\frac{M_{0}}{M_{\kappa}}} e_{\kappa\alpha,\nu}(\mathbf{q}) e^{i\mathbf{q}\cdot\mathbf{R}_{p}} e^{-i\omega_{\mathbf{q}\nu}t} ~,
\end{equation}
where the constants $u_{\mathbf{q}\nu}$ have to be determined by the initial conditions.
In the absence of external fields and in the thermodynamic limit,
these should yield a thermalized distribution of the atomic displacements.
For a finite supercell,
one could use, for example, the ZG displacements introduced in Refs.~\onlinecite{ZachariasPRB2016,ZachariasPRR2020}
as an initial configuration.
In a more refined treatment, these initial thermal displacements could be obtained starting
from a formulation of the problem using the Keldysh contour extended to include a vertical leg in the complex
plane \cite{Konstantinov1961,Stefanucci_Vanleeuwen_2013}. 
This non-equilibrium formulation
of the polaron problem is potentially promising and should be explored in future work.
%
A particular solution of Eq.~(\ref{eq:eqmo_disp_expec}) is given by:
\begin{equation} \label{eq:part_sol_ep}
	\langle \Delta \hat{\tau}_{\kappa\alpha p} \rangle
	= - \! \sum_{\kappa'\alpha' p'} C^{-1}_{\kappa\alpha p,\kappa'\alpha' p'}
	\int d\mathbf{r} \frac{\partial V_{\mathrm{tot}}^{0}(\mathbf{r})}{\partial \tau_{\kappa'\alpha' p'}} 
	\, 
    n_{\mathrm{e}}(\mathbf{r})~.
\end{equation}
The general solution of Eq.~(\ref{eq:eqmo_disp_expec}) is given by 
the sum of Eqs.~(\ref{eq:disp_homo_final}) and (\ref{eq:part_sol_ep}).
However, we note that the time-average of Eq.~(\ref{eq:disp_homo_final}) vanishes.
Furthermore, in the presence of anharmonic phonon-phonon couplings, which are not included 
in our Hamiltonian Eq.~(\ref{eq:elph_ham}), the amplitude of the fluctuations in 
Eq.~(\ref{eq:disp_homo_final}) must decay with a characteristic phonon lifetime.
Since we are interested in equilibrium properties of the polaronic state,
we neglect the fluctuations in Eq.~(\ref{eq:disp_homo_final}),
and focus on the static term in Eq.~(\ref{eq:part_sol_ep}) as the average of the atomic 
displacement operator.

The expectation value of the displacement operator in Eq.~(\ref{eq:part_sol_ep})
can be linked with the electron Green's function introduced in Sec.~\ref{sec:green}
via the standard relation:\cite{HedinLundqvistSSP1969}
\begin{align} \label{eq:n_from_G}
    \langle \hat{n}_{\mathrm{e}} (\mathbf{r}) \rangle 
    &= -i \hbar \, G(\mathbf{r}t,\mathbf{r}t^{+}) \nonumber \\
    &= -\frac{\hbar}{\pi} \int_{-\infty}^{\mu} \!\!\! d\omega \, \mathrm{Im}\left[ G(\mathbf{r},\mathbf{r};\omega) \right] ~,
\end{align}
where $\mu$ is the chemical potential. 
This relation derives directly from the definition of
$G$ in Eq.~(\ref{eq:Green_def_r}).

Combining Eqs.~(\ref{eq:Green_def_r})\FG{-}(\ref{eq:Green_def}), (\ref{eq:selfen_LP}), (\ref{eq:normal_mode_coord}), (\ref{eq:part_sol_ep}) and (\ref{eq:n_from_G}),
we find the following self-consistent expression for the polaronic self-energy:
\begin{multline} \label{eq:LP_from_G}
    \Sigma^{\mathrm{P}}_{n\mathbf{k},n'\mathbf{k'}}(t,t')
    =
    - \delta(t-t') 
    \, \frac{2}{N_p} \! \sum_{n'' n''' \mathbf{k''} \nu}
    \frac{g_{n n' \nu}(\mathbf{k}',\mathbf{k-k'}) }{\hbar\omega_{\mathbf{k-k'},\nu}} 
    \\
    \times \! \left[ -i\hbar \, G_{n'''\mathbf{k''+k-k'},n''\mathbf{k}''}(t,t^{+}) \right]
    \, g^{*}_{n''' n'' \nu}(\mathbf{k}'',\mathbf{k-k'}) ~.
\end{multline}
A diagrammatic representation of Eq.~(\ref{eq:LP_from_G}) is given in Fig.~\ref{fig:diagrams}~(c),
where we can recognize that the polaronic self-energy shows a tadpole structure 
similar to the Hartree self-energy term found in the electron-electron problem.\cite{Mattuck1992,MariniPRB2015}

Equations (\ref{eq:selfen_FM}), (\ref{eq:vertex_selfcon}) and (\ref{eq:LP_from_G}),
together with the Dyson equation given in Eq.~(\ref{eq:dyson_t}) and shown in Fig.~\ref{fig:diagrams}(d),
form a closed set of self-consistent equations,
similar to the Hedin equations in the electron-electron problem.\cite{HedinPR1965,HedinLundqvistSSP1969}
A graphical representation of the interdependence of the different elements is shown in Fig.~\ref{fig:diagrams}(e).

Calculations of polarons using these equations are beyond the reach of current computational methods. 
In the following sections we introduce standard approximations that make this problem tractable and 
amenable to \textit{ab initio} calculations.

\section{Many-body polaron equations}\label{sec:poleq}

\begin{figure}
    \includegraphics[width=0.85\columnwidth]{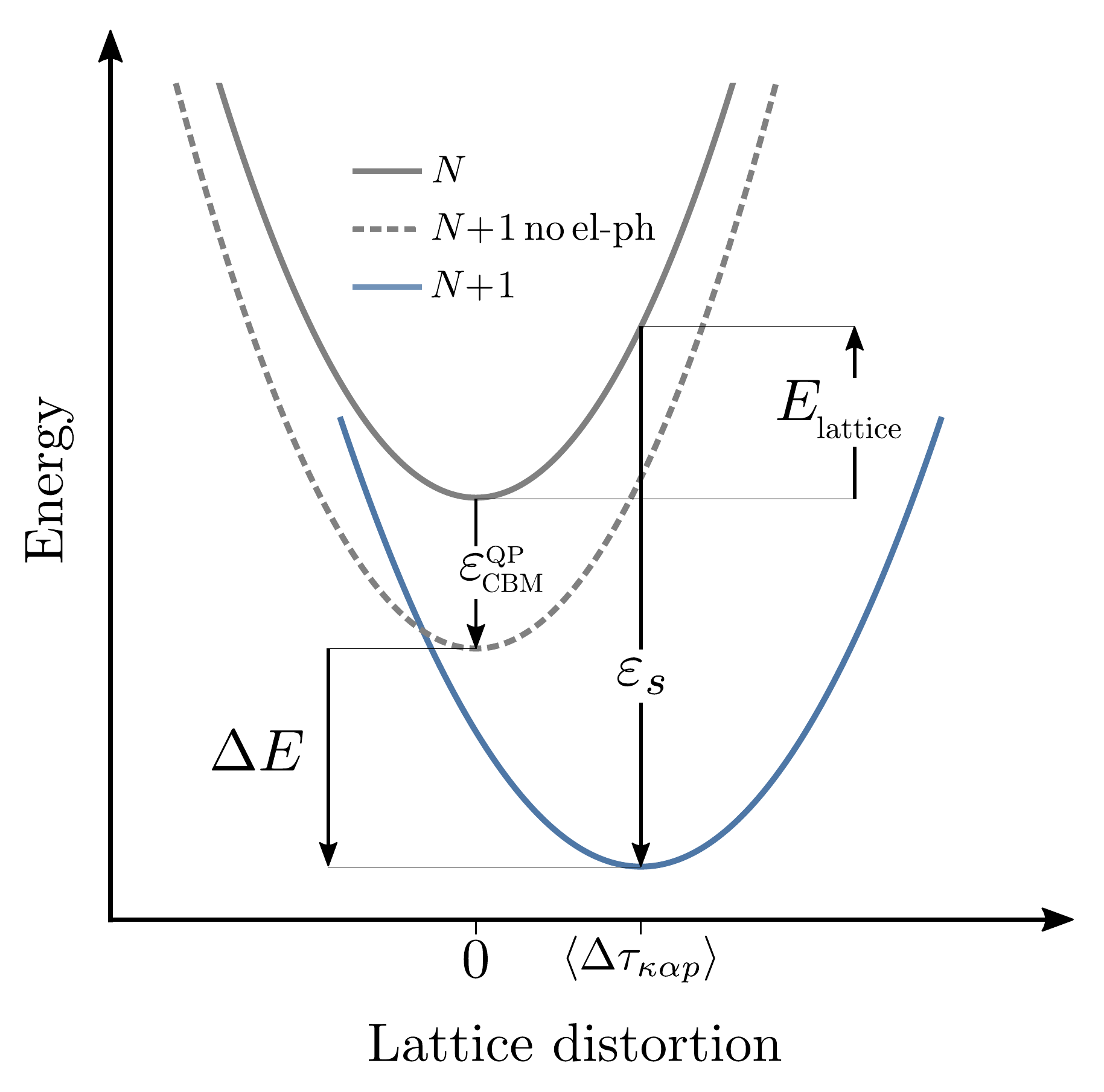}
    \caption{Schematic representation of the different energies involved in the polaron formation process.
    The solid gray line represents the original periodic $N$-electron system,
    and the solid blue line represents the $N+1$-electron system
    for which a distorted polaron configuration is the ground state in the presence of a pinning
    potential (cf.\ Sec.~\ref{sec:loc}).
    The dashed gray line represents a fictitious $N+1$ electron system in which 
    an electron with no phonon-mediated renormalization has been added to the conduction band minimum ($\varepsilon^{\mathrm{QP}}_{\mathrm{CBM}}$).
    The excitation energy from the $N+1$-electron ground state to the distorted $N$-electron state
    is the 
    electron addition/removal
    energy $\varepsilon_s$.
    The energy released by the relaxation of the lattice back to the periodic configuration is represented by $E_{\mathrm{lattice}}$.
    The polaron formation energy,
    that is the energy gained by the system when a delocalized electronic state becomes localized in
     a polaronic state,
    is represented by $\Delta E$.
    \label{fig:polaron_energy_graph}}
\end{figure}
%

\subsection{Lehmann representation in the polaron problem} \label{sec:lehmann}

In this section we move from the Green's function to 
Dyson orbitals
using the Lehmann representation.
To begin with, we recall that 
the Fourier transform of the Green's function can be written in the Lehmann representation 
as:\cite{FetterWalecka2003}
\begin{equation} \label{eq:G_Lehmann}
    G(\mathbf{r},\mathbf{r}';\omega)
    =
    \sum_{s} \frac{f_{s}(\mathbf{r}) f^{*}_{s}(\mathbf{r'})}{\hbar \omega - \left[ \varepsilon_{s} + i\eta\,\mathrm{sgn}(\mu - \varepsilon_{s}) \right]} ~,
\end{equation}
where $\eta\rightarrow 0^{+}$, the functions $f_s$ are Dyson orbitals, and the energies $\varepsilon_s$ are
electron addition/removal
energies. We define these quantities below.
Since our reference state is the $N\!+\!1$-electron system,
our notation for the Lehmann representation differs slightly from the conventional notation.
In Eq.~(\ref{eq:G_Lehmann}), the 
electron addition/removal
energies  are defined as:
\begin{eqnarray}
      \varepsilon_{s} &=& E_{N+2,s}-E_{N+1}  \hspace{10pt} \mathrm{for}\quad \varepsilon_{s}\geq \mu \label{eq:lehmann_e_1} ~, \\[4pt]
      \varepsilon_{s} &=& E_{N+1}-E_{N,s} \hspace{20pt}  \mathrm{for}\quad \varepsilon_{s}< \mu \label{eq:lehmann_e_2} ~,
\end{eqnarray}
where $E_{N+1\pm1,s}$ is the energy corresponding to the ${|N+1\pm1,s\rangle}$ many-body eigenstate
of the Hamiltonian in Eq.~(\ref{eq:elph_ham}), 
$E_{N+1}$ is the ground state energy of the reference $N\!+\!1$-particle system,
and $\mu=\frac{\partial E_{N+1}}{\partial N}$.
The Dyson orbitals $f_s$ are given by:\cite{HedinLundqvistSSP1969}
\begin{eqnarray} \label{eq:lehmann_f}
      f_{s}(\mathbf{r}) &=& \langle \, N+1 \,| \, \hat{\psi}(\mathbf{r}) \, | \,N+2,s \, \rangle   \hspace{10pt}\mathrm{for} \quad \varepsilon_{s}\geq \mu \label{eq:lehmann_f_1} ~, \\[4pt]
      f_{s}(\mathbf{r}) &=& \langle \, N,s \,| \, \hat{\psi}(\mathbf{r}) \, | \,N+1 \, \rangle \hspace{28pt}  \mathrm{for} \quad \varepsilon_{s}< \mu \label{eq:lehmann_f_2} ~.
\end{eqnarray}
The physical interpretation of these Dyson orbitals within the polaron problem is discussed below.
Given that we are mainly interested in occupied polaron states,
we will focus on the case $\varepsilon_{s}< \mu$.
A similar reasoning holds for the case $\varepsilon_{s}\geq \mu$.

In Eq.~(\ref{eq:lehmann_f_2}), $f_{s}(\mathbf{r})$ gives the probability amplitude for
the state $\hat{\psi}^{\dagger}(\mathbf{r})|N,s\rangle$ to be contained in the reference 
$| N+1 \rangle$ ground state. The state  $\hat{\psi}^{\dagger}(\mathbf{r})|N,s\rangle$
is a quantum state (not necessarily an eigenstate) with $N+1$ electrons, which is obtained
by adding an electron at position $\mathbf{r}$ to the excited eigenstate $|N,s\rangle$ of the $N$-electron system.

Now, the many-body eigenstates $| N+1 \rangle$ and $|N,s\rangle$ correspond to correlated electron-phonon states. 
Since $\hat{\psi}(\mathbf{r})$ acts purely on the electronic component of these states,
for $f_{s}(\mathbf{r})$ to be 
sizeable,
the phonon part of $|N,s\rangle$ has to 
overlap significantly with
the phonon part in the $| N+1 \rangle$ ground state.
Thus,
if $| N+1 \rangle$ corresponds to 
a localized polaron configuration in which the atoms are displaced with respect to the periodic lattice,
the first
electron addition/removal
energy $\varepsilon_{s}$ 
with a sizeable Dyson amplitude
corresponds to the removal energy of the extra electron while the ions remain frozen
in the distorted structure.
Other 
$|N,s\rangle$ states with different atomic configurations
still having vibrational wave functions non-orthogonal to $| N+1 \rangle$
will yield finite but exponentially small Dyson amplitudes,
and will not be considered in the following
(this is best seen by considering the coherent-state approximation
of Sec.~\ref{sec:toten}).
The first $|N,s\rangle$ state yielding a sizeable $f_s(\mathbf{r})$
differs from the ground state of the $N$-electron system due to the lattice distortion.
In order to reach the $N$-electron ground state,
the energy released by the distorted lattice upon relaxation must be accounted for.
A schematic representation of this process,
and its relation to the polaron formation energy (which will be discussed in Sec.~\ref{sec:toten}),
is given in Fig.~\ref{fig:polaron_energy_graph}.

The Dyson orbital for the lowest energy state of the $N$-particle system, say
$f_{s_{\rm min}}({\bf r})$, has a simple physical interpretation if we approximate the
exact many-body state $|N,s_{\rm min}\rangle$ by a single Slater determinant. 
Indeed, in this case,
all valence electronic states must be occupied, and $f_{s_{\mathrm{min}}}(\mathbf{r})$ constitutes the
lowest-energy single particle wavefunction in the conduction manifold. Therefore this Dyson orbital
represents the electronic part of the polaron wave function.

\subsection{Self-energies in terms of the polaron quasiparticle amplitudes} \label{sec:selfen_from_Ank}

Following Ref.~\onlinecite{SioPRB2019},
we proceed by expanding the Dyson orbitals in a single-particle basis:
\begin{equation} \label{eq:dyson_orb_expansion}
    f_{s}(\mathbf{r}) = N_{p}^{-1/2} \sum_{n\mathbf{k}} A_{n\mathbf{k}}^{s} \, \psi_{n\mathbf{k}}(\mathbf{r}) ~.
\end{equation}
In the following we refer to the coefficients $A_{n\mathbf{k}}^{s}$ as the polaron quasiparticle amplitudes.
\raggedbottom
After combining Eqs.~(\ref{eq:normal_mode_coord}), (\ref{eq:part_sol_ep}), (\ref{eq:n_from_G}), (\ref{eq:G_Lehmann}) and (\ref{eq:dyson_orb_expansion}),
and using the standard relations between the electron-phonon matrix elements,
interatomic force constant matrix,
and vibrational eigenmodes,\cite{GiustinoRMP2017,SioPRB2019}
we can express the expectation values of the normal mode coordinates as:
\begin{align} \label{eq:zqv_with_Anks}
    \langle \hat{z}_{\mathbf{q}\nu} \rangle
    =&
    -2 N_{p}^{-3/2} \sqrt{\frac{\hbar}{2M_{0}\omega_{\mathbf{q}\nu}}} \nonumber \\
    &\times \sum_{s}^{\varepsilon_s < \mu} \! \sum_{\mathbf{k}nn'} 
    \, A_{n'\mathbf{k+q}}^{s}
    \frac{g^{*}_{n'n\nu}(\mathbf{k},\mathbf{q})}{\hbar\omega_{\mathbf{q}\nu}}
    \, (A_{n\mathbf{k}}^{s})^{*}~.
\end{align}
Using this expression inside Eq.~(\ref{eq:selfen_LP}), and transforming to the frequency domain,
we obtain the polaronic self-energy in terms of the polaron quasiparticle
amplitudes:
%
    \begin{multline} \label{eq:Sigma_plrn}
        \Sigma^{\mathrm{P}}_{n\mathbf{k},n'\mathbf{k'}}(\omega) 
        = - \frac{2}{N_{p}^2} \sum_{\nu}
        \, g_{nn'\nu}(\mathbf{k'},\mathbf{k-k'}) \\
        \times\sum_{s}^{\varepsilon_s < \mu} \! \sum_{{\mathbf k}''mm'}
        \!\!\! 
        A_{m'\mathbf{k}''+\mathbf{k-k'}}^{s}
        \, \frac{g_{m'm\nu}^{*}(\mathbf{k}'',\mathbf{k-k'})}{\hbar\omega_{\mathbf{k-k'}\nu}}
        (A_{m\mathbf{k}''}^{s})^{*}
        \,.\,
    \end{multline}
\raggedbottom
We proceed similarly for the FM self-energy. In view of practical calculations, we approximate
the vertex function in Eq.~(\ref{eq:vertex_selfcon}) by 
retaining only the term in the first line.\cite{MigdalJETP1958,EngelsbergSchriefferPR1963}
This is the well-known Migdal approximation, and is a standard procedure in the electron-phonon literature.\cite{Grimvall1981,GiustinoRMP2017}
With this approximation, the FM self-energy in Eq.~(\ref{eq:selfen_FM}) becomes:
\begin{widetext}
\begin{equation} \label{eq:Migdal_selfen_offdiag}
    \Sigma^{\mathrm{FM}}_{n\mathbf{k},n'\mathbf{k'}}(t,t')
    =
    \frac{i}{N_{p}} \sum_{\substack{m \mathbf{k}'' \nu \\ m' \mathbf{k}''' \nu' }}
    g^{*}_{mn\nu}(\mathbf{k},\mathbf{k}''-\mathbf{k})
    \, g_{m'n'\nu'}(\mathbf{k}',\mathbf{k}'''-\mathbf{k'})
    \, G_{m\mathbf{k''},m'\mathbf{k}'''} (t,t')
    \, D_{\mathbf{k''}-\mathbf{k}\nu,\mathbf{k}'''-\mathbf{k'}\nu'}(t,t') ~,
\end{equation}
%
having used the relation
$g_{nn'\nu}(\mathbf{k'},\mathbf{k-k'})=g^{*}_{n'n\nu}(\mathbf{k},\mathbf{k}'-\mathbf{k})$ 
to achieve a compact expression. 
Along similar lines as in Eq.~(\ref{eq:Sigma_plrn}), 
we write the FM self-energy in the frequency domain and in terms of the quasiparticle amplitudes, by combining
Eqs.~(\ref{eq:G_Lehmann}), (\ref{eq:dyson_orb_expansion}), and (\ref{eq:Migdal_selfen_offdiag}):
%
\begin{equation} \label{eq:FM_selfen_Fourier2}
    \Sigma^{\mathrm{FM}}_{n\mathbf{k},n'\mathbf{k'}}(\omega)
    =
    \frac{i}{N_{p}^{2}} \!\!\! \sum_{\substack{n'' \mathbf{k}'' \nu \\ n''' \mathbf{k}''' \nu' }}
    g^{*}_{n''n\nu}(\mathbf{k},\mathbf{k}''-\mathbf{k})
    \, g_{n'''n'\nu'}(\mathbf{k}',\mathbf{k}'''-\mathbf{k'})
    \int\frac{d\omega'}{2\pi}
    \sum_{s} \frac{A_{n''\mathbf{k''}}^{s} \left ( A_{n'''\mathbf{k}'''}^{s} \right )^{*} D_{\mathbf{k''}-\mathbf{k}\nu,\mathbf{k}'''-\mathbf{k'}\nu'}(\omega') }{\hbar \omega - \hbar \omega' - \varepsilon_{s} - i\eta\,\mathrm{sgn}(\mu-\varepsilon_{s})}
    ~.
\end{equation}
In order to proceed further,
we approximate the interacting phonon Green's function by its 
adiabatic
counterpart:\cite{GiustinoRMP2017}
\begin{equation} \label{eq:D_approx}
    D^{0}_{\mathbf{q}\nu,\mathbf{q'}\nu'}(\omega)
    = \left[\frac{1}{\omega-\omega_{\mathbf{q}\nu}+i\eta} - \frac{1}{\omega+\omega_{\mathbf{q}\nu}-i\eta} \right]
    \delta_{\mathbf{q}\mathbf{q}'}\delta_{\nu\nu'} ~.
\end{equation}
This approximation is well justified and usually very accurate because the 
vibrational
frequencies are obtained from DFPT calculations. By inserting
Eq.~(\ref{eq:D_approx}) inside Eq.~(\ref{eq:FM_selfen_Fourier2}),
and performing the integral over $\omega''$ by closing the contour in the upper half of the complex plane,
we find:
%
    \begin{align} \label{eq:Sigma_FM_Fourier}
        \Sigma^{\mathrm{FM}}_{n\mathbf{k},n'\mathbf{k'}}(\omega)
        =&~
        \frac{1}{N_{p}^{2}} \sum_{mm' \mathbf{q} \nu}
        g^{*}_{mn\nu}(\mathbf{k},\mathbf{q})
        \, g_{m'n'\nu}(\mathbf{k}',\mathbf{q}) \nonumber \\
        &\times \sum_{s} A_{m\mathbf{k}+\mathbf{q}}^{s} \left ( A_{m'\mathbf{k'}+\mathbf{q}}^{s} \right )^{*}
        \left[ \frac{\theta(\varepsilon_{s}-\mu)}{\hbar \omega - \varepsilon_{s} - \hbar \omega_{\mathbf{q}\nu} + i\eta}
        + \frac{\theta(\mu-\varepsilon_{s})}{\hbar \omega - \varepsilon_{s} + \hbar \omega_{\mathbf{q}\nu} - i\eta} \right] ~,
    \end{align}
%
where $\theta$ is the Heaviside function.

\raggedbottom

\subsection{Self-consistent polaron equations} \label{sec:qp_eq}

The self-energies $\Sigma^{\rm P}$ and $\Sigma^{\rm FM}$ obtained in the previous section
can be used inside the Dyson equation Eq.~(\ref{eq:Dyson_inverted}) after a transformation to frequency domain:
\begin{equation} \label{eq:Dyson_w}
    G^{-1}_{n\mathbf{k},n'\mathbf{k'}}(\omega)
    =
    (G^{0})^{-1}_{n\mathbf{k},n'\mathbf{k'}}(\omega) - \Sigma^{\rm P}_{n\mathbf{k},n'\mathbf{k'}}(\omega)
    - \Sigma^{\rm FM}_{n\mathbf{k},n'\mathbf{k'}}(\omega) ~.
\end{equation}
The first term in the right-hand side is obtained from Eq.~(\ref{eq:nonint_G0})
by using the Fourier representation of the Dirac delta function:
\begin{equation} \label{eq:G0_Fourier}
    (G^{0})^{-1}_{n\mathbf{k},n'\mathbf{k'}}(\omega)
    =
    \left ( \hbar \omega - \varepsilon_{n\mathbf{k}} \right )
    \, \delta_{n\mathbf{k},n'\mathbf{k'}} ~.
\end{equation}
Note that $(G^{0})^{-1}_{n\mathbf{k},n'\mathbf{k'}}(\omega)$ in Eq.~\eqref{eq:G0_Fourier},
$\Sigma^{\mathrm{P}}_{n\mathbf{k},n'\mathbf{k'}}(\omega)$ in Eq.~\eqref{eq:Sigma_plrn},
and $\Sigma^{\mathrm{FM}}_{n\mathbf{k},n'\mathbf{k'}}(\omega)$ in Eq.~\eqref{eq:Sigma_FM_Fourier}
have dimensions of energy.

We are now in a position to combine the above results into a
self-consistent set of equations for the polaron quasiparticle amplitudes. By using
Eqs.~(\ref{eq:G_Lehmann}), (\ref{eq:dyson_orb_expansion}), (\ref{eq:Sigma_plrn}), 
and (\ref{eq:Sigma_FM_Fourier})-(\ref{eq:G0_Fourier}), we find that the poles of the
interacting Green's function are the solutions of the following eigenvalue problem:
        \begin{equation}\label{eq:qp_eq_general}
        \sum_{n'\mathbf{k}'} H^{\mathrm{pol}}_{n\mathbf{k},n'\mathbf{k'}} A_{n'\mathbf{k}'}^{s}
        = \varepsilon_{s} A_{n\mathbf{k}}^{s} ~,
        \end{equation}
where the effective polaron Hamiltonian $H^{\mathrm{pol}}$ depends on the 
electron addition/removal
energies
and quasiparticle amplitudes as follows:,
    \begin{eqnarray}
        &&\hspace{-10pt}H^{\mathrm{pol}}_{n\mathbf{k},n'\mathbf{k'}}
        =
        \varepsilon_{n\mathbf{k}} \delta_{n\mathbf{k},n'\mathbf{k'}}
        \left. - \, \frac{2}{N_{p}^{2}}
        \sum_{\substack{mm' \\ \nu {\mathbf k}''}} \sum_{s}^{\varepsilon_s < \mu}
        \, A_{m'\mathbf{k}''+\mathbf{k}-\mathbf{k}'}^{s}
        \, \frac{g^{*}_{m'm\nu}(\mathbf{k}'',\mathbf{k}-\mathbf{k'})}{\hbar\omega_{\mathbf{k}-\mathbf{k'}\nu}}
        \,(A_{m\mathbf{k}''}^{s})^{*}
        \, g_{nn'\nu}(\mathbf{k}',\mathbf{k}-\mathbf{k'}) \right. \nonumber \\
        &&\hspace{10pt}+ \frac{1}{N_{p}^{2}} \sum_{\substack{m m' \\ \nu \mathbf{q}}} \sum_{s'} 
        \, g^{*}_{m'n\nu}(\mathbf{k},\mathbf{q})
        \, g_{mn'\nu}(\mathbf{k}',\mathbf{q})
        A_{m'\mathbf{k}+\mathbf{q}}^{s'} ( A_{m\mathbf{k}'+\mathbf{q}}^{s'} )^{*}
        \! \left[ \frac{\theta(\varepsilon_{s'}-\mu)}{\varepsilon_{s} - \varepsilon_{s'} - \hbar \omega_{\mathbf{q}\nu} + i\eta}
        + \frac{\theta(\mu-\varepsilon_{s'})}{\varepsilon_{s} - \varepsilon_{s'} + \hbar \omega_{\mathbf{q}\nu} - i\eta}
         \right].
        \label{eq:qp_eq_general2}
    \end{eqnarray}
\end{widetext}
The self-consistent solution of Eqs.~(\ref{eq:qp_eq_general})-\eqref{eq:qp_eq_general2} yields the excitation energies 
and the quasiparticle amplitudes of the Dyson orbitals, and hence the polaron energies and wavefunctions.
We note that the only approximations that we have made thus far are
the Migdal approximation to the electron-phonon vertex in Eq.~(\ref{eq:vertex_selfcon}), 
and the replacement of the interacting phonon Green's function by its noninteracting (i.e. DFPT) 
counterpart in Eq.~(\ref{eq:phonon_Green1}).

Equations~\eqref{eq:qp_eq_general}-\eqref{eq:qp_eq_general2} constitute the central result of this manuscript. These equations
generalize the polaron equations derived in Refs.~\onlinecite{SioPRL2019,SioPRB2019} within the context
of DFPT to a many-body Green's function formalism for the polaron quasiparticle amplitudes and
excitation energies. Practical strategies for solving these equations are outlined in Sec.~\ref{sec:abinitio}.

\begin{figure}
    \includegraphics[width=0.6\columnwidth]{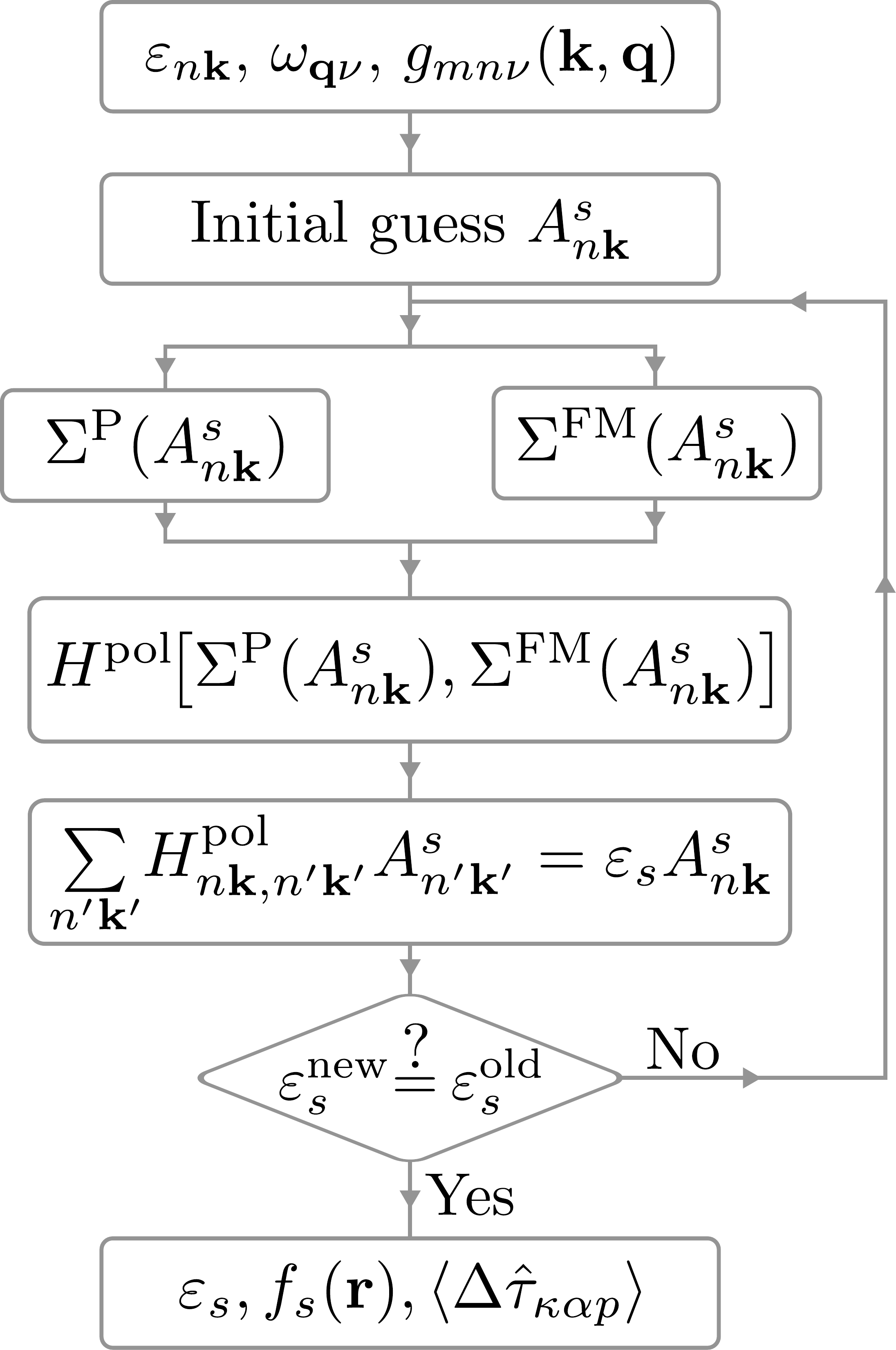}
    \caption{Schematic representation of the self-consistent procedure required to solve the many-body polaron equations, Eqs.~(\ref{eq:qp_eq_general}) and \eqref{eq:qp_eq_general2}.
    \label{fig:qp_eq_flow}}
\end{figure}

A schematic illustration of the self-consistent procedure required for solving 
Eqs.~(\ref{eq:qp_eq_general})-\eqref{eq:qp_eq_general2} is provided in Fig.~\ref{fig:qp_eq_flow}.

%
%

\subsection{Total energy of the polaron ground state} \label{sec:toten}

In this section,
we derive an expression for the total energy of the polaronic ground state in terms of the 
Dyson orbitals and the eigenvalues of Eq.~(\ref{eq:qp_eq_general}).
We proceed along the same lines as for the Galitskii-Migdal formula,\cite{GalitskiiMigdalJETP1958,HolmPRB2000}
except that we consider the coupled
electron-phonon Hamiltonian in Eq.~(\ref{eq:elph_ham}).

Upon acting on Eq.~(\ref{eq:eqmo_cnk}) with $\hat{c}^{\dagger}_{n'\mathbf{k'}}(t')$
and taking the expectation value over the ground state, we find:
\begin{eqnarray} \label{eq:eqmo_cnk_2}
    &&i\hbar \frac{\partial}{\partial t} \langle \hat{c}^{\dagger}_{n'\mathbf{k'}}(t') \hat{c}_{n\mathbf{k}}(t) \rangle
    = \varepsilon_{n\mathbf{k}} \langle \hat{c}^{\dagger}_{n'\mathbf{k'}}(t') \hat{c}_{n\mathbf{k}}(t) \rangle 
    + N_p^{-\frac{1}{2}} \times
     \nonumber \\
    &&\times \sum_{n' \mathbf{q} \nu } g_{nn'\nu}(\mathbf{k}\!-\!\mathbf{q},\mathbf{q})
    \langle \hat{c}^{\dagger}_{n'\mathbf{k'}}(t') \hat{c}_{n'\mathbf{k-q}}(t)
    ( \hat{a}_{\mathbf{q}\nu}+\hat{a}_{-\mathbf{q}\nu}^\dagger ) \rangle.\hspace{-1cm}\nonumber\\ 
\end{eqnarray}
This expression can be related to the total energy, $E=\langle \hat{H} \rangle$, by taking
the expectation value of the electron-phonon Hamiltonian in Eq.~(\ref{eq:elph_ham}):
\begin{eqnarray}
    E &=& 
    \lim_{\substack{
    t'\rightarrow t^{+} \\ 
    n'\mathbf{k'}= n\mathbf{k}}} 
    \sum_{n\mathbf{k}} i\hbar \frac{\partial}{\partial t} \langle \hat{c}^{\dagger}_{n'\mathbf{k'}}(t') \hat{c}_{n\mathbf{k}}(t) \rangle \nonumber \\ &+&
\sum_{\mathbf{q}\nu} \hbar\omega_{\mathbf{q}\nu} (\langle \hat{a}_{\mathbf{q}\nu}^\dagger \hat{a}_{\mathbf{q}\nu} \rangle+1/2).
\end{eqnarray}
The first term on the right hand side of this expression can be identified with the electron Green's function from Eq.~(\ref{eq:Green_def}):
\begin{eqnarray} \label{eq:GM_elph}
    E &=& 
    \lim_{\substack{t'\rightarrow t^{+} \nonumber \\ 
    n'\mathbf{k'}= n\mathbf{k}}} 
    \sum_{n\mathbf{k}} \hbar^{2} \frac{\partial}{\partial t} G_{n\mathbf{k},n'\mathbf{k'}} (t,t')  \\
    & +& \sum_{\mathbf{q}\nu} \hbar\omega_{\mathbf{q}\nu} (\langle \hat{a}_{\mathbf{q}\nu}^\dagger \hat{a}_{\mathbf{q}\nu} \rangle+1/2) ~.
\end{eqnarray}
We transform this result into the frequency domain, and we make
use of the spectral representation of the Green's function:\cite{HolmPRB2000}
\begin{multline} \label{eq:GM_elph_Fourier}
    E = - \sum_{n\mathbf{k}} \frac{\hbar^{2}}{\pi}
    \int_{-\infty}^{\mu} \! d\omega \, \omega \, \mathrm{Im}\left[ G_{n\mathbf{k},n\mathbf{k}}(\omega) \right] \\
    + \sum_{\mathbf{q}\nu} \hbar\omega_{\mathbf{q}\nu} (\langle \hat{a}_{\mathbf{q}\nu}^\dagger \hat{a}_{\mathbf{q}\nu} \rangle+1/2) ~.
\end{multline}
By further using
the Lehmann representation in Eq.~(\ref{eq:G_Lehmann}) and
the expansion of the Dyson orbitals in terms of polaron quasiparticle amplitudes, 
Eq.~(\ref{eq:dyson_orb_expansion}), we obtain:
\begin{align} \label{eq:toten_gen}
    E = \frac{1}{N_{p}} \sum_{n\mathbf{k}}
    \sum_{s}^{\varepsilon_s < \mu} \varepsilon_{s} \, |A_{n\mathbf{k}}^{s}|^2 
    + \sum_{\mathbf{q}\nu} \hbar\omega_{\mathbf{q}\nu} (\langle \hat{a}_{\mathbf{q}\nu}^\dagger \hat{a}_{\mathbf{q}\nu} \rangle+1/2) ~.
\end{align}
In the second term on the right hand side, the expectation value of the phonon number operator
must also be related to the polaron quasiparticle amplitudes $A_{n\mathbf{k}}^{s}$. We have not found a way
to establish this relation in the most general case. However, an accurate first-principles formulation
is still possible if we make the approximation that the phonon subsystem can be described as a superposition of
coherent states. 

Coherent states are minimum-uncertainty wavepackets, and in the case of
the harmonic oscillator they corresponds to Gaussian wavefunctions, rigidly translated away from the minimum
of the potential well. The reason for considering coherent states is that much of the earlier
literature on the Fr\"ohlich polaron model shows how coherent states constitute a very accurate variational
ansatz for determining the ground state energy of the polaron.\cite{Allcock1956,Hohler1954,Bogolyubov2016,Buimistrov1957a,Buimistrov1957b,Berezin1986}

We therefore approximate the phonon subsystem of the polaron ground state as the following normalized 
superposition of coherent states:
  \begin{equation} \label{eq:coh_ansatz}
    |N+1\rangle \!=\! \exp \! \left[\sum_{\mathbf{q}\nu} \left( u_{\mathbf{q}\nu} 
        \hat a^\dagger_{\mathbf{q}\nu} \!-\! |u_{\mathbf{q}\nu}|^2/2 \right) \right]
        \!|N+1,0_{\rm ph}\rangle,
  \end{equation}
where $u_{\mathbf{q}\nu}$ indicates the (complex) displacement of the wavepacket and, $|0_{\rm ph}\rangle$
denotes the phonon vacuum. With this approximation, we have the standard property 
$\hat a_{\mathbf{q}\nu}|N+1\rangle = u_{\mathbf{q}\nu} |N+1\rangle$. 
By using this relation inside Eqs.~\eqref{eq:toten_gen}, \eqref{eq:zqnu}, and 
\eqref{eq:zqv_with_Anks}, and employing time-reversal symmetry to replace $u^*_{-\mathbf{q}\nu}$ by $u_{\mathbf{q}\nu}$,
we can rewrite the total energy as:
\begin{equation} \label{eq:toten_gen2}
    E = \frac{1}{N_{p}} \sum_{n\mathbf{k}}
    \sum_{s}^{\varepsilon_s < \mu} \varepsilon_{s} \, |A_{n\mathbf{k}}^{s}|^2 
    + \sum_{\mathbf{q}\nu} \hbar\omega_{\mathbf{q}\nu} (|u_{\mathbf{q}\nu}|^2+1/2) ~,
\end{equation}
where the coherent displacements are given by:
\begin{equation} \label{eq:cohdisp}
    u_{\mathbf{q}\nu}
    =
    -N_{p}^{-3/2}  \sum_{s}^{\varepsilon_s < \mu} \! \sum_{\mathbf{k}nn'}
    A_{n'\mathbf{k+q}}^{s}
    \, \frac{g_{n'n\nu}^{*}(\mathbf{k},\mathbf{q})}{\hbar\omega_{\mathbf{q}\nu}}
    (A_{n\mathbf{k}}^{s})^{*}~.
\end{equation}
The last two equations provide the relation between
the total ground state energy of an interacting electron-phonon system, and
the excitation energies and quasiparticle amplitudes of the Dyson orbitals, within the
approximation of coherent states for the phonon subsystem. 
From the second term on the right hand side of Eq.~\eqref{eq:toten_gen2}, 
we see that $|u_{\mathbf{q}\nu}|^2$
gives the number of phonons $\mathbf{q}\nu$ (per BvK supercell) 
contributing to 
the polaronic lattice distortion.

\section{Toward \textit{ab initio} calculations} \label{sec:abinitio}

\subsection{Approximations for practical calculations}\label{sec:approx}

The formalism developed in Sec.~\ref{sec:poleq} provides a self-consistent mathematical framework 
to investigate polaron wave functions and formation energies within a first-principles many-body
approach.
However, the self-consistent solution of Eqs.~(\ref{eq:qp_eq_general}) and \eqref{eq:qp_eq_general2} 
is currently beyond reach for real materials, because it requires a
summation over all the occupied and unoccupied polaronic states in the Fan-Migdal self-energy term,
including those with finite total polaron momentum.

In view of devising a practical approach for systematic many-body \textit{ab initio} calculations of polarons,
we make the following reasoning. If we approximate the interacting many-body ground state by a single Slater determinant,
and assume that the added electron in the $(N+1)$-electron system has a negligible effect on the
lowest $N$ electron wavefunctions and energies, the contributions of the valence states in the
$N$- and $(N+1)$-electron systems in Eq.~(\ref{eq:toten_gen}) are identical.
This approximation is physically motivated by the fact that the addition of a single electron to a system of many electrons will modify the electron density only slightly 
\cite{SioPRB2019}. 
Furthermore, by construction, the $N$-electron system is associated with a periodic undistorted lattice,
therefore the expectation value of the phonon number operator vanishes in $\langle N | \hat{H} | N \rangle$.
These observations lead to the following simplified expression for the formation energy of the polaron 
in the BvK supercell
\footnote{The equivalent expression for the formation energy of a hole polaron
upon electron removal is
\mbox{$\Delta E = -(\varepsilon_{s,\mathrm{min}} - \varepsilon^{\mathrm{QP}}_{\mathrm{VBM}}) + \sum_{\mathbf{q}\nu} \hbar \omega_{\mathbf{q}\nu} |u_{\mathbf{q}\nu}|^{2}$.}}
:
\begin{align} \label{eq:eform_2}
    \Delta E
    &= \langle N+1 | \hat{H} | N+1 \rangle 
    - \left( \langle N | \hat{H} | N \rangle + \varepsilon^{\rm QP}_{\mathrm{CBM}} \right) \nonumber \\
    &= \varepsilon_{s,\mathrm{min}} - \varepsilon^{\rm QP}_{\mathrm{CBM}} 
    + \sum_{\mathbf{q}\nu} \hbar\omega_{\mathbf{q}\nu} |u_{\mathbf{q}\nu}|^2~.
\end{align}
In this expression, $\varepsilon^{\rm QP}_{\mathrm{CBM}}$ represents the many-body electron addition energy
for the periodic, undistorted lattice. For example, this could be the quasiparticle energy of the
conduction band bottom in a GW calculation in the absence of atomic displacements.
A schematic illustration of Eq.~(\ref{eq:eform_2}) is given in Fig.~\ref{fig:polaron_energy_graph}.
The quantity $E_{\rm lattice}$ appearing in the figure corresponds to the last term of 
Eq.~\eqref{eq:eform_2}, where the coherent displacements are given by Eq.~\eqref{eq:cohdisp}.

\raggedbottom

Similarly, we obtain a compact expression for the polaronic self-energy $\Sigma^{\rm P}$ 
by replacing the first $N$ occupied states with unperturbed Bloch wavefunctions
With this choice, the quasiparticle amplitudes for these states become Dirac delta functions, 
$A_{n\mathbf{k}}^{s}\!=\!\sqrt{N_p}\,\delta_{s,n\mathbf{k}}$, and their contribution 
to $\Sigma^{\rm P}$
can be neglected in Eq.~(\ref{eq:Sigma_plrn}).
As a result, only the lowest-energy Dyson orbital, 
which corresponds to the electronic part of the polaron wave function (cf.\ Sec.~\ref{sec:lehmann}),
contributes in the summation in Eq.~(\ref{eq:Sigma_plrn}):
\begin{multline} \label{eq:LP_simple}
    \Sigma^{\mathrm{P}}_{n\mathbf{k},n'\mathbf{k'}} 
    = - \frac{2}{N_{p}^{2}}
    \sum_{\substack{mm' \nu {\mathbf k}'' }}
    g_{nn'\nu}(\mathbf{k'},\mathbf{k-k'}) \\
    \times A_{m'\mathbf{k}''+\mathbf{k-k'}}^{s,\mathrm{min}}
    \, \frac{g_{m'm\nu}^{*}(\mathbf{k}'',\mathbf{k-k'})}{\hbar\omega_{\mathbf{k-k'}\nu}}
    \, (A_{m\mathbf{k}''}^{s,\mathrm{min}})^{*}
\end{multline}
A similar reasoning can be extended to the self-energy $\Sigma^{\rm FM}$ in Eq.~\eqref{eq:Migdal_selfen_offdiag}.
We replace the interacting Green's
function $G$ by its non-interacting counterpart $G^{0}$.
This choice amounts to assuming sharp quasiparticles, as in the
standard G$_0$W$_0$ approximation.\cite{HybertsenLouiePRB1986}
With this replacement, $\Sigma^{\rm FM}$
becomes diagonal in the single-particle basis, and simplifies to:
\begin{eqnarray} \label{eq:FM_diag_G0}
    &&\Sigma^{\mathrm{FM}}_{n\mathbf{k},n'\mathbf{k'}}(\omega)
    = \frac{\delta_{n\mathbf{k},n'\mathbf{k'}}}{N_{p}} \sum_{m \mathbf{q} \nu}
    |g_{mn\nu}(\mathbf{k},\mathbf{q})|^2 \times \nonumber \\
    &&
    \left[ \frac{\theta(\varepsilon_{m \mathbf{k+q}}-\mu)}{\hbar\omega - \varepsilon_{m \mathbf{k+q}} - \hbar\omega_{\mathbf{q}\nu} + i\eta}
    + \frac{\theta(\mu-\varepsilon_{m \mathbf{k+q}})}{\hbar\omega - \varepsilon_{m \mathbf{k+q}} + \hbar\omega_{\mathbf{q}\nu} - i\eta} \right]. \nonumber \\
 \end{eqnarray}
This expression
only depends on the phonon frequencies and the electron-phonon matrix elements of the periodic configuration.
Equation~(\ref{eq:FM_diag_G0}) is the standard expression used in 
\textit{ab initio} calculations of electron-phonon renormalization of band structures.\cite{GiustinoRMP2017}
%
%

Using the above simplifications, the many-body polaron equations are reduced to solving
the self-consistent eigenvalue problem given by
\begin{multline} \label{eq:qp_eq_simplified_1}
    \sum_{n'\mathbf{k}'} \Big \{
    \varepsilon_{n\mathbf{k}} \delta_{n\mathbf{k},n'\mathbf{k}'}
    + \Sigma^{\mathrm{P}}_{n\mathbf{k},n'\mathbf{k'}} \\
    + \Sigma^{\mathrm{FM}}_{n\mathbf{k},n'\mathbf{k}'}(\varepsilon_{s,\mathrm{min}}/\hbar)
    \Big \}
    \, A^{s,\mathrm{min}}_{n'\mathbf{k}'} = \varepsilon_{s,\mathrm{min}} \, A^{s,\mathrm{min}}_{n\mathbf{k}} ~,
\end{multline}
where the self-energies are given by Eqs.~(\ref{eq:LP_simple}) and (\ref{eq:FM_diag_G0}). In the
remainder of this manuscript, the index $s_{\mathrm{min}}$ corresponding to the wavefunction and
eigenvalue of the polaron in the ground state will been omitted for ease of notation.

\subsection{Relation to the theory of Ref.~\onlinecite{SioPRB2019}} \label{sec:relation_to_Sio}

The many-body polaron equations Eqs.~(\ref{eq:LP_simple})\FG{-}(\ref{eq:qp_eq_simplified_1})
share a similar form with the DFPT polaron equations derived in 
Ref.~\onlinecite{SioPRB2019}, see Eqs.~(37) and (38) of that work.
In fact,
if we neglect the Fan-Migdal self-energy term in Eq.~\eqref{eq:qp_eq_simplified_1}, 
the two sets of equations become identical within the approximations outlined in Sec.~\ref{sec:approx}.

The main differences between the two approaches are that, 
in the present case, 
(i) the polaron eigenvalue also incorporates the dynamical Fan-Migdal self-energy renormalization,
and (ii) the lattice distortion energy in Eq.~\eqref{eq:toten_gen} is directly linked to
the number of phonons that participate in the polaron via the phonon number operator 
$\hat{a}_{\mathbf{q}\nu}^\dagger \hat{a}_{\mathbf{q}\nu}$.

The fact that we reached a very similar set of polaron equations as in Ref.~\onlinecite{SioPRB2019}
starting from a general many-body formulation is very encouraging, and provides a rigorous
field-theoretic justification for the DFPT approach followed in Ref.~\onlinecite{SioPRB2019}.

At a qualitative level, the main improvement of the present many-body approach over the DFPT
strategy of Ref.~\onlinecite{SioPRB2019} is in that our present formalism incorporates dynamical effects
as described by the FM self-energy. 
Therefore, in addition to the physics of phonon-induced localization and self-trapping, 
the present approach also captures the physics of phonon-induced band 
structure renormalization and polaron satellites in photoemission spectra.

\subsection{Perturbation theory on the polaron quasiparticle amplitudes and energies} \label{sec:pert}

Equations~(\ref{eq:LP_simple})-(\ref{eq:qp_eq_simplified_1}) constitute a nonlinear, self-consistent
eigenvalue problem. The polaron energy $\varepsilon$ appears on both sides of Eq.~\eqref{eq:qp_eq_simplified_1},
therefore an iterative solution is required. 

This requirement can be relaxed if we proceed to evaluate the equations in perturbation theory. 
Specifically, one could solve the equations by retaining only $\Sigma^{\rm P}$ and treating $\Sigma^{\rm FM}$ 
within perturbation theory, or viceversa by retaining $\Sigma^{\rm FM}$ and treating
$\Sigma^{\rm P}$ perturbatively. 
Since $\Sigma^{\rm FM}$ does not couple different wavevectors, the latter
option would lead to a vanishing polaronic correction and no localization, 
which is equivalent to standard calculations of band
renormalization in absence of polarons. 
Therefore we focus on the former option of retaining
only the polaronic self-energy and treating the FM term perturbatively.
This procedure can be implemented in two steps:
\begin{enumerate}
    \item[(i)] Solve Eq.~(\ref{eq:qp_eq_simplified_1}) by 
    considering only the polaronic self-energy $\Sigma^{\rm P}$:
    \begin{equation} \label{eq:LP_only_qp}
      \hspace{0.8cm}  \sum_{n'\mathbf{k}'} \left\{
        \varepsilon_{n\mathbf{k}} \delta_{n\mathbf{k},n'\mathbf{k}'}
        + \Sigma^{\mathrm{P}}_{n\mathbf{k},n'\mathbf{k'}} \right\}
        A^{\mathrm{P}}_{n'\mathbf{k}'} = \varepsilon^{\mathrm{P}} A^{\mathrm{P}}_{n\mathbf{k}} ~.
    \end{equation}
    \item[(ii)] Add the FM contribution to the polaron energy after replacing
         $A_{n\mathbf{k}}$ by the solution at the previous step, $A^{\mathrm{P}}_{n\mathbf{k}}$:
\begin{eqnarray} \label{eq:epsilon_from_Ank}
    \hspace{1cm} \varepsilon &=& \frac{1}{N_p}\!\sum_{n\mathbf{k}}
    \sum_{n'\mathbf{k}'} \! A^{{\rm P},*}_{n\mathbf{k}} \!
    \left(\varepsilon_{n\mathbf{k}} \delta_{n\mathbf{k},n'\mathbf{k}'}
    \!+\!\Sigma^{\mathrm{P}}_{n\mathbf{k},n'\mathbf{k'}} \right) \!
    A^{\rm P}_{n'\mathbf{k}'}
    \nonumber \\
    &&+ \frac{1}{N_p}\sum_{n\mathbf{k}} |A^{\mathrm{P}}_{n\mathbf{k}}|^{2} \, 
    \Sigma^{\mathrm{FM}}_{n\mathbf{k}}(\omega).
\end{eqnarray}
\end{enumerate}
Within the simplest Rayleigh-Schr\"odinger perturbation theory, the frequency $\omega$ appearing in the last equation can either be set the polaron eigenvalue, $\omega = \varepsilon^{\rm P}/\hbar$, or to the unperturbed Bloch eigenvalue, $\omega = \varepsilon^0/\hbar$. In Sec.~\ref{sec:frohlich} 
we compare the latter to the self-consistent solution of Eq.~\eqref{eq:qp_eq_simplified_1}.

%

\section{Applications} \label{sec:appl}

\subsection{The Fr{\"o}hlich model}\label{sec:frohlich}

To validate the theory developed in Secs.~\ref{sec:theory}-\ref{sec:abinitio},
we apply the formalism to the Fr\"ohlich model.\cite{LandauPZS1933,PekarZETF1946,FrohlichPHM1950}
The Fr\"ohlich model represents a standard benchmark in the study of polaron physics, and has been investigated by a number
of authors using a variety of many-body techniques.\cite{LeeLowPinesPRB1953,FrohlichADP1954,FeynmanPR1955,
ProkofevPRL1998,MischenkoPRB2000,GrusdtPRB2016,Devreese2020arXiv}. The availability of highly-accurate solutions
such as Feynman's path integral results\cite{FeynmanPR1955} and diagrammatic Monte Carlo calculations\cite{MischenkoPRB2000}
makes it possible to carefully assess the validity of our approach and of
the approximations described in Sec.~\ref{sec:abinitio}.

\begin{figure}[t]
    \includegraphics[width=1.0\columnwidth]{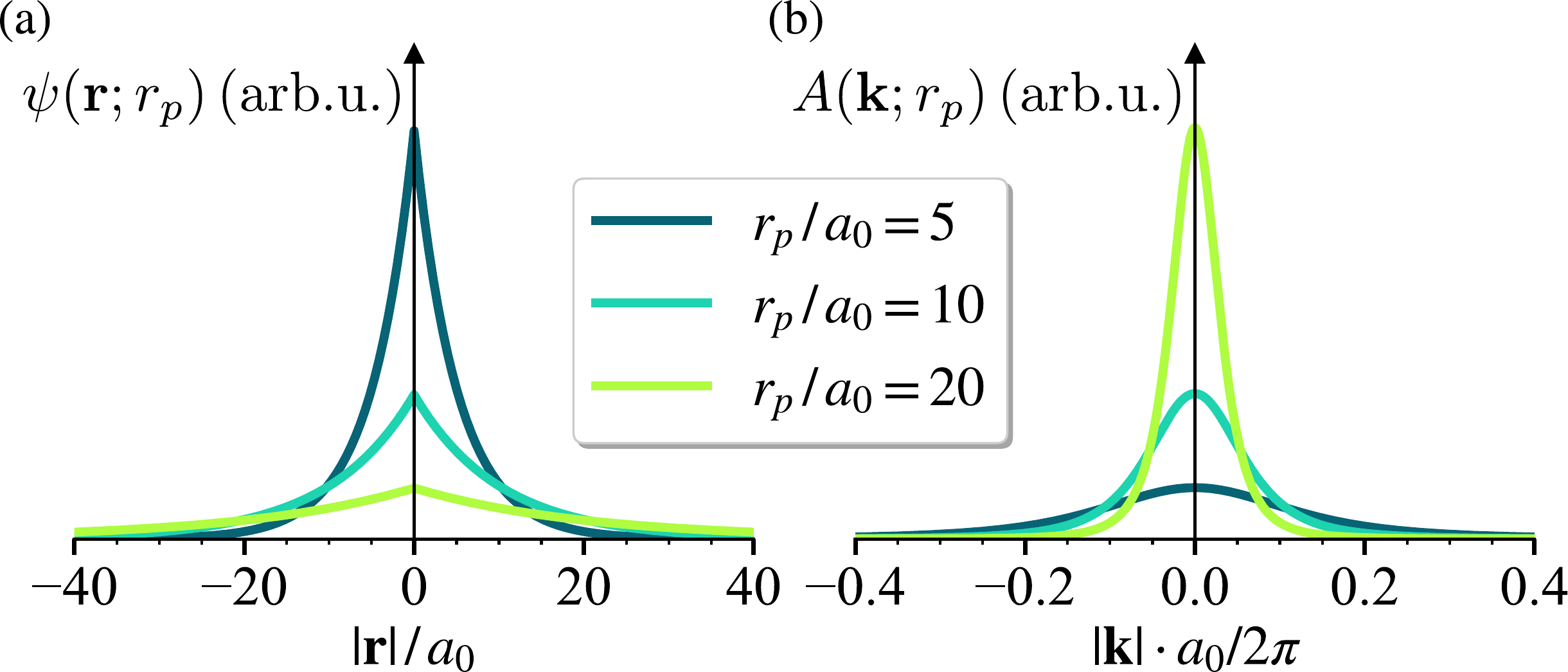}
    \caption{Variational ansatz for the electronic component of the polaron wave function in the Fr{\"o}hlich model.
     (a) Wavefunction plot for a few values of the variational parameter $r_p$, which corresponds to the polaron
      radius, from Eq.~(\ref{eq:polaron_wf_r}). $a_0$ is the Bohr radius. (b) 
      Reciprocal-space coefficients of the wavefunctions shown in (a), from Eq.~(\ref{eq:polaron_wf_k}).
    \label{fig:psir_ak}}
\end{figure}

In the Fr{\"o}hlich model, 
the Hamiltonian given by Eq.~(\ref{eq:elph_ham}) is simplified by considering
a single electron band with effective mass $m^{*}$
and parabolic dispersions $\varepsilon_{\mathbf{k}} = \hbar^{2}|\mathbf{k}|^2/2m^*$,
coupled to a dispersionless longitudinal polar optical phonon with frequency $\omega_{\mathrm{LO}}$.
The coupling matrix element is given by:
\cite{FrohlichADP1954,VerdiPRL2015}
\begin{equation} \label{eq:matel_frohlich}
    g(\mathbf{q}) = \frac{i}{|\mathbf{q}|} \, \left[\frac{e^2}{4\pi\epsilon_{0}}\frac{4\pi}{\Omega}\frac{\hbar\omega_{\mathrm{LO}}}{2} \frac{1}{\kappa} \right]^{1/2} ~.
\end{equation}
In this equation,
$\epsilon_{0}$ is the vacuum permittivity, and the dielectric screening constant $\kappa$ is defined by
$1/\kappa = 1/\epsilon^{\infty}-1/\epsilon^{0}$, with $\epsilon^{\infty}$ and $\epsilon^{0}$ being
the high-frequency electronic permittivity and the static dielectric constant including the ionic contribution,
respectively. 
In this model,
the Debye-Waller self-energy vanishes \cite{KandolfPRB2022},
and the electron-phonon coupling strength is traditionally described by a single parameter $\alpha$,
referred to as the Fr{\"o}hlich coupling constant:\cite{Mahan1993,Alexandrov2010,SioPRB2019}
\begin{equation} \label{eq:Frohlich_alpha}
    \alpha = \frac{e^2}{4\pi\epsilon_{0}} \frac{1}{\hbar} \sqrt{\frac{m^{*}}{2\hbar\omega_{\mathrm{LO}}}} \frac{1}{\kappa} ~.
\end{equation}
There is a single Dyson orbital,
which we identify with the electronic part of the polaron wave function, $f(\mathbf{r})=\psi(\mathbf{r})$.
The expansion in Eq.~(\ref{eq:dyson_orb_expansion})
can now be performed in terms of plane waves,
and the transition to the extended
crystal is performed by considering an infinite number of unit cells in the BvK supercell
so that summations over the momentum $\mathbf{k}$ become continuous integrals:
\begin{equation} \label{eq:Fourier_cont}
    \psi(\mathbf{r})= \frac{\sqrt{\Omega}}{(2\pi)^3} \int d\mathbf{k} ~ A(\mathbf{k}) ~ e^{i\mathbf{k}\cdot \mathbf{r}}~.
\end{equation}
Here, $\mathbf{r}$ and $\mathbf{k}$ belong to $\mathbb{R}^3$,
and $\Omega$ is the unit cell volume.
We require that
the polaron wave function be normalized in real space,
\begin{equation} \label{eq:normalization_r}
    \int d\mathbf{r} ~ |\psi(\mathbf{r})|^2 = 1 ~,
\end{equation}
and this implies the normalization of its Fourier coefficients:
\begin{equation} \label{eq:normalization_k}
    \frac{\Omega}{(2\pi)^3} \int d\mathbf{k} ~ |A(\mathbf{k})|^2 = 1 ~.
\end{equation}
Using Eqs.~\eqref{eq:Fourier_cont} and (\ref{eq:normalization_k}) inside Eq.~(\ref{eq:qp_eq_simplified_1}), we obtain:
\begin{multline} \label{eq:qp_Frohlich}
    \varepsilon
    = \frac{\Omega}{(2\pi)^3} \int d\mathbf{k} \, A(\mathbf{k})
    \int d\mathbf{k'}
    \bigg[ 
        \varepsilon_{\mathbf{k}} \, \delta(\mathbf{k-k'}) \\
        + \Sigma^{\mathrm{P}}(\mathbf{k},\mathbf{k'})
        + \Sigma^{\mathrm{FM}}(\mathbf{k};\varepsilon) \, \delta(\mathbf{k-k'})
    \bigg]
    A(\mathbf{k'}) ~.
\end{multline}
Using Eq.~(\ref{eq:LP_simple}), the polaronic self-energy appearing in this expression becomes:
\begin{equation} \label{eq:LP_Frohlich}
    \Sigma^{\mathrm{P}}(\mathbf{k},\mathbf{k'})
    =
    -\frac{2\Omega^{2}}{(2\pi)^6}
    \, \frac{|g(\mathbf{k-k'})|^2}{\hbar\omega_{\mathrm{LO}}}
    \int d\mathbf{k}''
    A_{\mathbf{k''+k-k'}} A^*_{\mathbf{k}''} ~.
\end{equation}
The FM self-energy in Eq.~(\ref{eq:FM_diag_G0}) can be evaluated exactly,\cite{Mahan1993} 
and is given by:
\begin{equation} \label{eq:FM_frohlich_analytical}
    \Sigma^{\rm FM}(\mathbf{k}; \varepsilon) = - \frac{\alpha \, (\hbar \omega_{\mathrm{LO}})^{3/2}}{\sqrt{\varepsilon_{\mathbf{k}}}} 
    \arcsin \left (\sqrt{\frac{\varepsilon_{\mathbf{k}}}{\hbar\omega_{\mathrm{LO}} - \varepsilon + \varepsilon_{\mathbf{k}}}} \right ) ~.
\end{equation}
To calculate the energy and wave function of the lowest polaron state, we use a variational approach.
For simplicity,
following Refs.~\onlinecite{Alexandrov2010,SioPRB2019}, 
for the electronic part
we employ a normalized exponential trial wave function:
\begin{equation} \label{eq:polaron_wf_r}
    \psi(\mathbf{r};r_p)= \sqrt{\frac{1}{\pi r_{p}^3}} \exp\left[-|\mathbf{r}|/r_{p}\right] ~,
\end{equation}
where $r_p$ can be identified as the polaron radius.
The Fourier transform of this function is:
\begin{equation} \label{eq:polaron_wf_k}
    A(\mathbf{k};r_p) =
8
\sqrt{\frac{\pi r_{p}^3}{\Omega}}
\frac{1}{(r_p^2 |\mathbf{k}|^2 + 1)^2} ~.
\end{equation}
%
Equations~(\ref{eq:polaron_wf_r}) and (\ref{eq:polaron_wf_k})
show that the more localized the wave function is in real space (small $r_p$),
the more extended are its coefficients in Fourier space,
and vice-versa.
In Fig.~\ref{fig:psir_ak}, 
we show the exponential trial wave function for different values of the polaron radius $r_p$,
together with the corresponding Fourier transforms.

We now use Eq.~(\ref{eq:polaron_wf_k}) in Eq.~(\ref{eq:qp_Frohlich}), and evaluate the integrals
for each of the three terms within square brackets. The first term is the average of the kinetic energy:
\begin{equation} \label{eq:expec_kin}
    \frac{\Omega}{(2\pi)^3} \! \int d\mathbf{k} \, A^{*}(\mathbf{k};r_p) \frac{\hbar|\mathbf{k}|^2}{2m^*} A(\mathbf{k};r_p)
    = \frac{\hbar}{2 m^* r_p^2} ~,
\end{equation}
and is identical to what is found in the Landau-Pekar model.\cite{Alexandrov2010,SioPRB2019}
The expectation value of the term containing $\Sigma^{\rm P}$ in Eq.~(\ref{eq:qp_Frohlich}) corresponds
to the Coulomb energy in the Landau-Pekar model, and is given by:\cite{Alexandrov2010,SioPRB2019}
\begin{equation} \label{eq:expec_LP_final}
    \langle \Sigma^{\mathrm{P}} \rangle 
    = -\frac{e^2}{4\pi\epsilon_{0}} \frac{1}{\kappa} \frac{5}{8}\frac{1}{r_p} ~.
\end{equation}
From Eqs.~(\ref{eq:qp_Frohlich}) and (\ref{eq:polaron_wf_k}), 
we see that the expectation value of the FM self-energy results from the radial integral:
\begin{equation} \label{eq:expec_FM_Frohlich}
    \langle \Sigma^{\mathrm{FM}} \rangle
    = \frac{\Omega}{(2\pi)^3}
    \,4\pi \!\! \int_{0}^{\infty} \!\!\! dk \, |A(k;r_p)|^2 \, \Sigma^{\rm FM}\left(k\right) ~.
\end{equation}
Let us analyze the asymptotic limits of this integral.
In the limit of a strongly localized polaron ($r_p\rightarrow 0$),
$A(k)$ tends to a constant value, but $\Sigma^{\rm FM}(k)$ is significant only
near $k=0$ [cf.\ Eq.~(\ref{eq:FM_frohlich_analytical})]. Owing to the normalization
of the Fourier coefficients, the integral vanishes in this limit. In the limit
of an extended polaron ($r_p\rightarrow \infty$),
the $A(k)$ coefficients become a Dirac delta function centered at $k=0$,
therefore the integral coincides with the value of the FM self-energy at $k=0$,
$\langle \Sigma^{\mathrm{FM}} \rangle=-\alpha\hbar\omega_{\mathrm{LO}}$ (having set $\varepsilon=0$).

\begin{figure*}
    \includegraphics[width=0.85\linewidth]{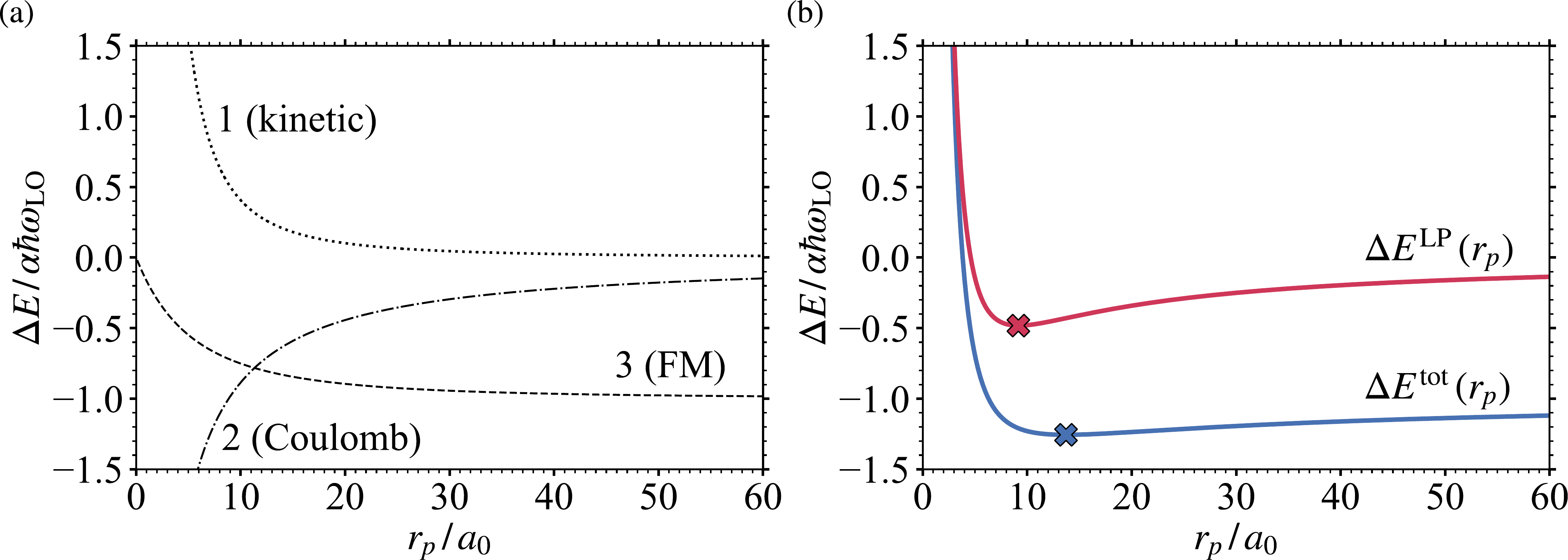}
    \caption{Ground state polaron energy as a function of the polaron radius $r_p$, 
            within the Fr{\"o}hlich model for $\alpha=4.94$.
            (a) $r_p$-dependence of the different terms contributing to the total energy in Eq.~(\ref{eq:E_vs_rp}),
            namely the kinetic energy (dotted),
            the Coulomb energy composed by the Landau-Pekar and the lattice energy (dotted-dashed),
            and the Fan-Migdal self-energy contribution (dashed).
            (b) Total energy as a function of the polaron radius $r_p$. 
            The red solid line represents the Landau-Pekar result,\cite{SioPRB2019} where only the kinetic and the Coulomb energies are considered.
            The blue solid line represents the total energy including the FM contribution, as in Eq.~(\ref{eq:E_vs_rp}).
            The energy minima are highlighted by the solid crosses.
    \label{fig:E_vs_rp}}
\end{figure*}

Equations~(\ref{eq:expec_kin})-(\ref{eq:expec_FM_Frohlich}) allow us to evaluate the
polaron eigenvalue as a function of the polaron radius, $\varepsilon(r_p)$. To determine the total
formation
energy, we also need to consider the lattice relaxation energy, i.e. the last term in Eq.~\eqref{eq:eform_2}.
The evaluation of this term yields:
\begin{multline} \label{eq:expec_phonon_1}
    \int d\mathbf{q} \, \hbar\omega_{\mathbf{q}} \langle \hat{a}^{\dagger}_{\mathbf{q}} \hat{a}_{\mathbf{q}} \rangle
    =
    \frac{\Omega^{3}}{(2\pi)^9} \int d\mathbf{q}
    ~\frac{|g(\mathbf{q})|^2}{\hbar\omega_{\mathrm{LO}}} \\
    \hspace{10pt}\times \int \! d\mathbf{k} \int \! d\mathbf{k}' 
    \, (A_{\mathbf{k}+\mathbf{q}})^* A_{\mathbf{k}} 
    \, A_{\mathbf{k'}+\mathbf{q}} (A_{\mathbf{k}'})^* 
    =
    \frac{e^2}{4\pi\epsilon_{0}} \frac{1}{\kappa} \frac{5}{16}\frac{1}{r_p} ~.\\[-20pt]
\end{multline}
Putting together the above results, we obtain the following expression for the total 
energy
of the Fr\"ohlich polaron as a function of the radius $r_p$:
\begin{equation} \label{eq:E_vs_rp}
    \Delta E(r_p) = \frac{\hbar}{2 m^* r_p^2} - \frac{5}{16}\frac{e^2}{4\pi\epsilon_{0}\kappa} \frac{1}{r_p}
            + \frac{2^8 \pi^2 r_{p}^3}{\Omega} \!\!\int_{0}^{\infty} \!\!\! dk \frac{ \Sigma^{\rm FM}(k) }{(1+r_p^2 k^2 )^4},
\end{equation}
where $\Sigma^{\rm FM}$ is given by Eq.~\eqref{eq:FM_frohlich_analytical}.
This total energy coincides with the energy of the Landau-Pekar model if we neglect the integral on the
right hand side [see for example Eq.~(10) of Ref.~\onlinecite{SioPRB2019}].
We note that the polaron formation energy within the Fr{\"o}hlich model 
has been called $E$ in previous work 
because the delocalized state with no electron-phonon interaction has zero total energy by definition.

In Fig.~\ref{fig:E_vs_rp} we analyze the total energy
as a function of the polaron radius $r_p$.
In this example, the physical parameters have been chosen to match those for
the electron polaron in LiF,\cite{SioPRB2019}
namely $m^{*}=0.88~m_{e}$, 
$\hbar \omega_{\mathrm{LO}}=77~\mathrm{meV}$,
$\epsilon^{0} = 10.62$ and $\epsilon^{\infty}=2.04$,
giving a coupling constant of $\alpha=4.94$.
Figure~\ref{fig:E_vs_rp}(a) illustrates the contribution to the polaron energy from each term in
Eq.~(\ref{eq:E_vs_rp}).
The first term on the right hand side of Eq.~(\ref{eq:E_vs_rp}) is the kinetic energy (dotted line). This term
is positive  and thus favors delocalization. The second term on the right hand side of Eq.~(\ref{eq:E_vs_rp}) 
is the Coulomb attraction energy between the electron and the lattice distortion (dashed-dotted line). This term
is negative and thus favors localization. The last term in Eq.~(\ref{eq:E_vs_rp}) is the FM self-energy
contribution. It is negative and thus it also favors localization, but it varies more smoothly with the
radius. As discussed above, this term tends to vanish at small radius, 
and approaches the value
$-\alpha \hbar\omega_{\rm LO}$ at large radius.

Figure \ref{fig:E_vs_rp}(b) shows the dependence of the total energy of the polaron on the radius (blue line).
The minimum of this energy is marked by a cross and indicates the variational solution.
For the sake of comparison, 
we also show the total energy curve for the Landau-Pekar model (red line).\cite{SioPRB2019} In this model there
is no FM contribution.
We see that the FM contribution modifies the shape of the energy surface of the Landau-Pekar model, and shifts
the minimum towards a larger radius and a lower ground state energy.

\begin{figure*}
    \includegraphics[width=0.875\linewidth]{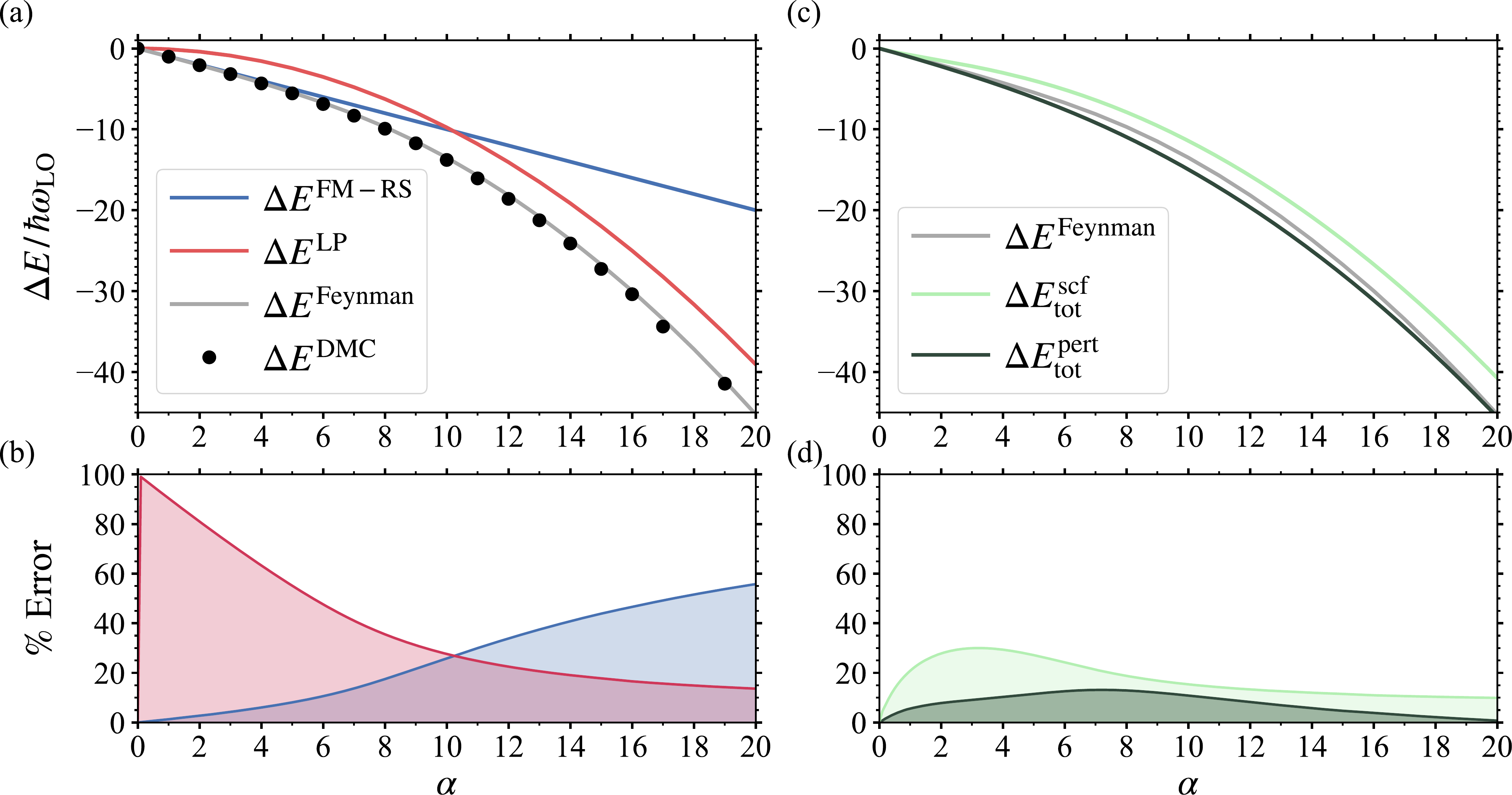}
    \caption{Total energy of the ground state of the Fr{\"o}hlich polaron as a function of the coupling strength $\alpha$.
    (a) Fan-Migdal solution in Rayleigh-Schr\"odinger perturbation theory (FM-RS, blue line),
    Landau-Pekar solution (LP, red line),
    Feynman's variational path-integral solution (gray line),\cite{FeynmanPR1955,SchultzPR1959,RosenfelderPLA2001}
    and diagrammatic Monte Carlo results (DMC, black circles).\cite{ProkofevPRL1998,MischenkoPRB2000,HahnPRB2018}
    \mbox{(b) Relative} deviation between FM-RS and LP energies with respect to Feynman's result,
    following the same color scheme as in (a).
    \mbox{(c) Ground-state} energy of the Fr{\"o}hlich polaron evaluated using the present 
      Green's function approach.
    The light green line represents the self-consistent solution, the
    dark-green line represents the perturbative calculation.
    The relative errors of each approximation with respect to Feynman's result are shown in (d).
    \label{fig:E_vs_alpha}}
\end{figure*}

Now we analyze the dependence of the variational polaron energy on the coupling constant $\alpha$.
To this aim, we generate curves like those in Fig.~\ref{fig:E_vs_rp}(b) for a range of parameters $\alpha$,
and we determine the minimum in each case. The results are reported in Fig.~\ref{fig:E_vs_alpha}.

In Fig.~\ref{fig:E_vs_alpha}(a),
the red line represents the Landau-Pekar ground state polaron formation energy.
Within the exponential ansatz used in Eq.~\eqref{eq:polaron_wf_r}, this energy is given by
$\Delta E^{\mathrm{LP}}=-(50/512)\alpha^{2}\hbar\omega_{\mathrm{LO}}$.\cite{SioPRB2019}
The blue line represents the FM self-energy evaluated within Rayleigh-Schr\"odinger perturbation theory,
and is given by $\Delta E^{\mathrm{FM,RS}}=-\alpha\hbar\omega_{\mathrm{LO}}$.\cite{Mahan1993}
The gray line represents the result obtained 
by Feynman's path integral method,\cite{FeynmanPR1955,SchultzPR1959,RosenfelderPLA2001}
and the black circles are diagrammatic Monte Carlo results.\cite{ProkofevPRL1998,MischenkoPRB2000,HahnPRB2018} 
This comparison shows that
Feynman's results are essentially as accurate as the diagrammatic Monte Carlo data, therefore
in the following we use Feynman's result as the ``exact'' solution for the purpose of comparison.

In Fig.~\ref{fig:E_vs_alpha}(b) we show the relative errors of the Landau-Pekar (LP) energy and 
the Fan-Migdal energy (in the Rayleigh-Schr\"odinger approximation, FM-RS)
with respect to Feynman's result
as filled areas, following the same color convention as in Fig.~\ref{fig:E_vs_alpha}(a).
It is clear that both approaches deviate significantly from Feynman's result throughout the
entire coupling range, with errors in the energy as large as 100\%.
The LP result gives the correct trend at strong couplings,
but it underestimates the polaron energy at weak couplings.
In contrast,
the FM-RS result correctly captures the linear dependence of the energy at weak couplings,
but it underestimates the polaron energy at strong couplings.

In Figs.~\ref{fig:E_vs_alpha}(c) and (d) we compare the total energy of the polaron calculated
using Eq.~\eqref{eq:E_vs_rp} with the Feynman theory. We compare two different levels of approximation. First,
we compute the polaron ground state energy by requiring self-consistency in the energy entering the FM self-energy, 
as in Eq.~(\ref{eq:qp_eq_simplified_1}).
We will denote the ground state energy obtained in this way by 
$\Delta E_{\mathrm{tot}}^{\mathrm{scf}}$.
Second,
we consider the perturbative approach discussed in Sec.~\ref{sec:pert}.
Within the Fr{\"o}hlich model,
this translates to the following two-step process for each coupling constant $\alpha$:
\begin{enumerate}
    \item[(i)] We determine the polaron radius that minimizes the Landau-Pekar total energy.
      With the exponential ansatz of Eq.~\eqref{eq:polaron_wf_r}, this radius is
    ${r_{p,{\rm min}} \!=\! 16 \kappa m_e a_0/ 5 m^*}$.\cite{SioPRB2019}
    \item[(ii)] We calculate the total formation energy of the polaron by
       adding the FM contribution evaluated at the non-interacting energy as a perturbation:
    \begin{multline}
       \hspace{20pt} \Delta E_{\mathrm{tot}}^{\mathrm{pert}} = -\frac{50}{512}\alpha^{2}\hbar\omega_{\mathrm{LO}} \\
        + 4\pi \int_{0}^{\infty} \! dk \, |A(k;r_{p,\mathrm{min}})|^2 \, \Sigma^{\rm FM}\left(k; \varepsilon=0\right) 
    \end{multline}
\end{enumerate}
The result of these two approaches are shown in 
Fig.~\ref{fig:E_vs_alpha}(c) as 
light green and dark green lines, respectively. The results by Feynman are shown as the dashed gray line.
The filled areas in Fig.~\ref{fig:E_vs_alpha}(d) represent the relative errors with respect to Feynman's result,
with the same color code as in Fig.~\ref{fig:E_vs_alpha}(c).
This comparison indicates that our formalism correctly describes the polaron energy throughout the entire
range of couplings, irrespective of the level of approximation adopted in the evaluation of the ground-state
energy.
Interestingly,
the deviation of the perturbative approach with respect to Feynman's results never exceeds $10\%$.
This success suggests that the perturbative procedure is particularly suitable for studying polarons,
and can be generalized to \textit{ab initio} calculations.

The variational ansatz employed in Eq.~\ref{eq:polaron_wf_r} could be improved further,\cite{PekarZETF1946,MiyakeJPSJ1975,Alexandrov2010}
therefore we expect that with some refinements we should be able to achieve an even better agreement 
with Feynman's theory. Since no other theoretical approach has succeeded to match Feynman's calculations 
at all couplings,\cite{Devreese2020arXiv} 
the present results are very encouraging, especially because the present
approach can be used for \textit{ab initio} calculations of real materials, as we show in Sec.~\ref{sec:LiF}.

\subsection{\textit{Ab initio} calculations in LiF} \label{sec:LiF}

\begin{figure}
    \includegraphics[width=0.8\columnwidth]{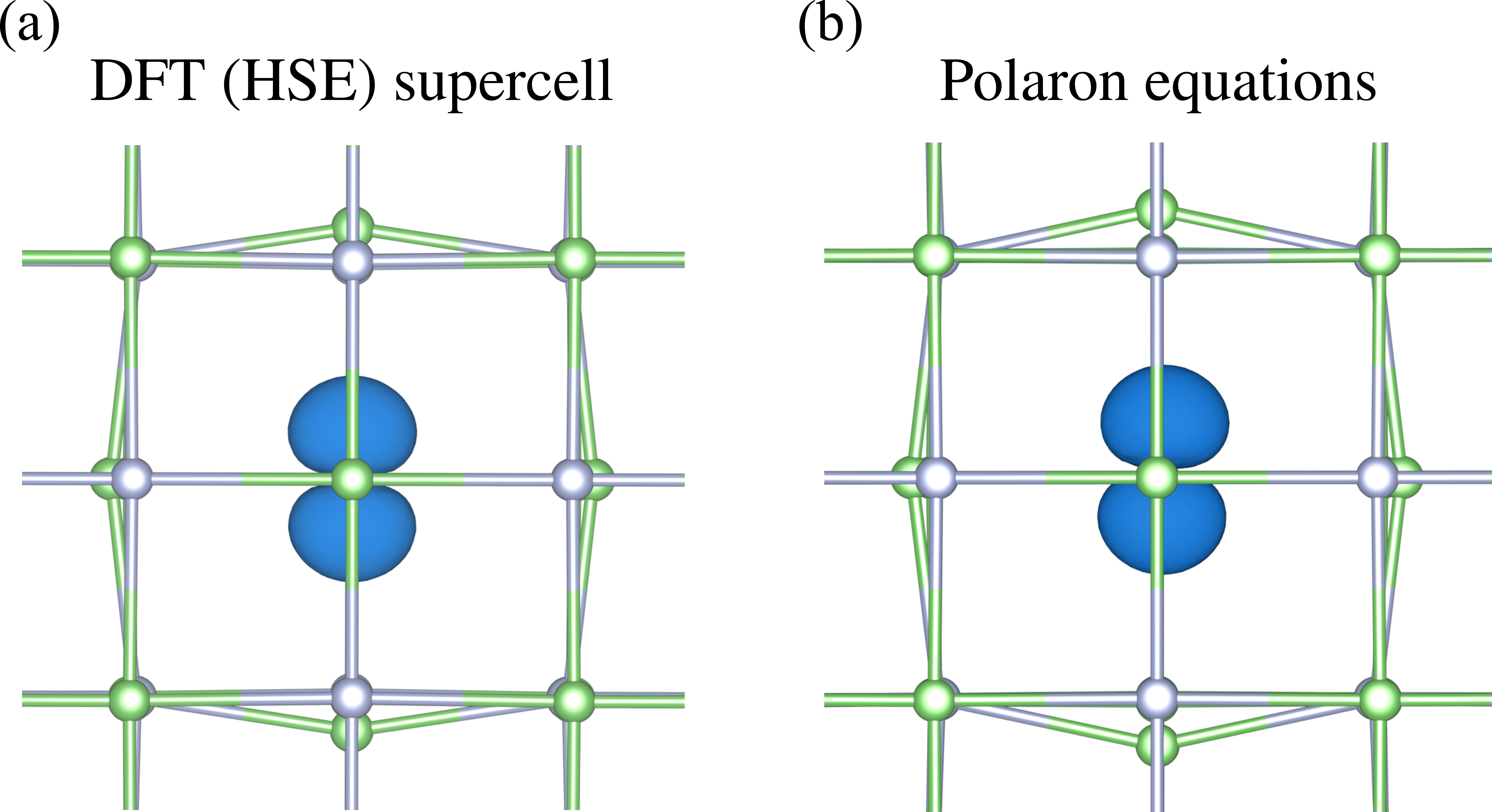}
    \caption{Comparison between the hole polaron wave function in LiF 
    obtained by
    (a) direct DFT (HSE) supercell calculation with a removed electron
    and (b) the solution of the polaron equations 
    in Eqs.~(\ref{eq:LP_simple}) and (\ref{eq:LP_only_qp}).
    \label{fig:HSE_vs_EPW}}
\end{figure}
\begin{figure*}[t]
    \includegraphics[width=1.0\linewidth]{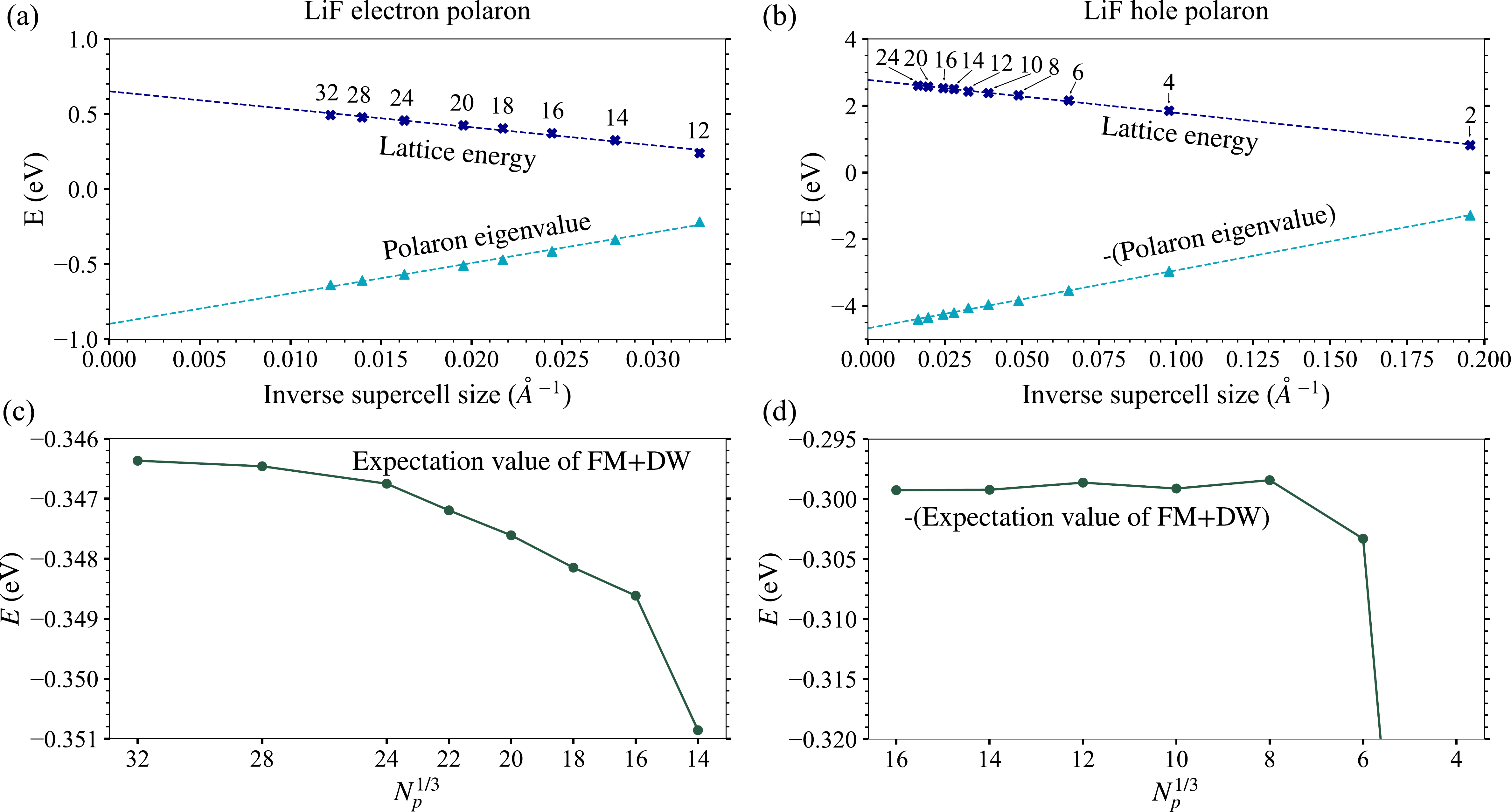}
    \caption{Analysis of the various contributions to the polaron formation energy in the lowest-order approximation, 
    Eqs.~(\ref{eq:eform_2}), (\ref{eq:LP_only_qp}) and (\ref{eq:epsilon_from_Ank}).
    (a) Polaron eigenvalue (light-blue triangles) and lattice energy (dark-blue crosses) as a function of the supercell size for the electron polaron in LiF.
    The supercell size is given as $L^{-1}$, where $L^{3}$ is the supercell volume.
    The numbers next to the data points indicate the number of unit cells in each direction of the homogeneous Born-Von Karman supercell,
    or equivalently the number of $\mathbf{k}$-points in each direction of the homogeneous mesh of the Brillouin zone.
    The dashed lines correspond to the extrapolation of each contribution to the infinite supercell size.
    (b) Same as (a) but for the hole polaron in LiF.
    (c) Expectation value of the FM+DW self-energy over the polaron quasiparticle amplitudes for the electron polaron in LiF,
    as a function of the number of $\mathbf{k}$-points in each direction of the homogeneous mesh of the Brillouin zone.
    (d) Same as (c) but for the hole polaron in LiF.
    \label{fig:LiF}}
\end{figure*}

As a first fully \textit{ab initio} calculation using the methodology presented in this work,
we compute the polaronic band gap renormalization of LiF from first principles.
To this aim, we consider the simplest approximation to our theory, as described in Sec.~\ref{sec:pert}.
In this section we provide the details of the computational procedure, while
the main results and implications are discussed in the companion manuscript.\cite{AccompanyingPaper}

All calculations are performed using the {\sc Quantum ESPRESSO} software suite.\cite{QE2017}
Ground state DFT calculations are performed within the 
Perdew-Burke-Ernzerhof generalized gradient approximation,\cite{PBEPRL1996}
using optimized norm-conserving Vanderbilt (ONCV) pseudopotentials \cite{HamannPRB2013,PseudoDojo}
and plane waves with a kinetic energy cutoff of 100 Ry.
Our optimized lattice parameter is $a=4.06$ \AA.
Phonon frequencies and electron-phonon matrix elements 
are computed within density functional perturbation theory.\cite{BaroniRMP2001}
Coarse momentum grids of 12$\times$12$\times$12 $\mathbf{k}$ 
and $\mathbf{q}$-points are used
for the ground state electron and lattice dynamics calculations, respectively.
Electron energies, phonon frequencies, and electron-phonon matrix elements
are interpolated to dense grids 
by means of Wannier-Fourier interpolation,\cite{MarzariPRB1997,SouzaPRB2001,GiustinoPRB2007}
as implemented in the wannier90 \cite{Wannier902020} and EPW \cite{EPW2016} codes.
The method presented in Ref.~\onlinecite{VerdiPRL2015} is used to deal with 
the long-range part of the electron-phonon matrix element for polar materials.

In order to converge both the band and momentum sums needed to compute the FM self-energy in Eq.~(\ref{eq:FM_diag_G0})
we proceed as follows.
Following Refs.~\onlinecite{GonzeAP2011,LihmPRB2020},
we divide the band sum into two subspaces: 
(i) a lower subspace formed by the valence band manifold 
and the first four conduction bands ($\sim 15$ eV above the conduction band bottom), 
where the momentum integration is carried out explicitely,
and (ii) an upper subspace formed by the rest of the conduction bands,
where the phonon frequency in the denominator of Eq.~(\ref{eq:FM_diag_G0}) is neglected 
and the band summation is transformed into the solution of a Sternheimer equation.\cite{GonzeAP2011}
For the solution of the Sternheimer equation and the calculation of the upper subspace contribution,
we employ the implementation of Ref.~\onlinecite{LihmPRB2020} 
within the {\tt PHonon} code.
The lower subspace contribution is calculated with EPW.
A slightly modified version of the code is used to evaluate the FM self-energy 
at the noninteracting polaron energy for all $\mathbf{k}$-points,
which corresponds to the Kohn-Sham energies of the band extrema.
The momentum integrals are converged by interpolating all quantities into fine 
96$\times$96$\times$96 $\mathbf{q}$-point grids for each 
$\Sigma^{\mathrm{FM}}_{\mathbf{k}}$.

Equations~(\ref{eq:LP_simple}) and (\ref{eq:LP_only_qp}) are solved iteratively
using the implementation of Refs.~\onlinecite{SioPRL2019,SioPRB2019} within EPW.
We initialize the $A_{n\mathbf{k}}$ coefficients 
using a Gaussian line shape centered at the band edge.
We note that,
similar to a DFT optimization,
different initializations could potentially lead to 
multiple local minima.
We validate the robustness of our results
by starting the iterative procedure 
with different Gaussian widths, 
as well as with random distributions of the $A_{n\mathbf{k}}$ coefficients,
for which equivalent self-consistent solutions are obtained within the convergence threshold
in all cases.
The lattice energy is evaluated using Eqs.~\eqref{eq:cohdisp}-\eqref{eq:eform_2}.
We use increasingly denser $\bf k$-point grids,
and we take the isolated polaron limit by extrapolating 
to infinite supercell size (infinitely dense $\mathbf{k}$-point grid).\cite{SioPRB2019}

To asses the validity of the approximation
that the change in the total density is negligible upon electron addition/removal
(see Sec.~\ref{sec:approx})
in the worst-case scenario,
we perform direct DFT calculations
on a $3\times 3\times 3$ supercell of LiF 
with an electron removed (hole polaron).
To mitigate the self-interaction error,
we use the Heyd-Scuseria-Ernzerhof (HSE)\cite{HSE2003} hybrid functional
with an exact exchange fraction parameter of $\alpha_{\mathrm{EXX}}=0.37$.
The atomic positions 
are relaxed so that forces on each ion are below $10^{-5}$ Ry/bohr.
Figure~\ref{fig:HSE_vs_EPW}(a) shows the relaxed atomic configuration,
together with the wave function of the first unuccupied Kohn-Sham state.
The total energy of the relaxed configuration
is lower than the the total energy obtained for the original periodic configuration
with the hole,
confirming that the polaronic configuration is more stable.
For comparison,
in Fig.\ref{fig:HSE_vs_EPW}~(b) we show the hole polaron wave function and atomic displacements obtained 
by the solution of Eqs.~(\ref{eq:LP_simple}) and (\ref{eq:LP_only_qp}),
which is practically identical to the result shown in Fig.~\ref{fig:HSE_vs_EPW}(a).
This result validates \textit{a posteriori} our initial assumption.

Figures~\ref{fig:LiF} (a) and (b) show our results for 
the electron and hole polaron energies in LiF, 
respectively. 
The light-blue triangles represent the polaron eigenvalues 
$\varepsilon^{\mathrm{P}}$ 
as a function of the inverse supercell size,
and dark-blue crosses represent the corresponding lattice energy in each case.
The dashed lines are used to extrapolate these quantities to infinite supercell size.
We obtain 
$\varepsilon^{\mathrm{P}}-\varepsilon^{\mathrm{KS}}_{CBM}=-0.898~\mathrm{eV}$ and 
$E_{\mathrm{lattice}}=0.652~\mathrm{eV}$ for the electron polaron,
and
$\varepsilon^{\mathrm{P}}-\varepsilon^{\mathrm{KS}}_{VBM}=4.672~\mathrm{eV}$ and
$E_{\mathrm{lattice}}=2.775~\mathrm{eV}$ for the hole polaron,
in good agreement with the results reported in Ref.~\onlinecite{SioPRB2019}.

To obtain the total polaron eigenvalue $\varepsilon$ from Eq.~(\ref{eq:epsilon_from_Ank})
we need to evaluate the average of the FM self-energy over the quasiparticle amplitudes $A_{n{\bf k}}$
obtained above.
We accomplish this by performing a Wannier interpolation of the FM self-energy, 
similar to the Wannier interpolation of the GW self-energy corrections to the band structure.\cite{HamannPRB2009}
This procedure consists of five steps:
(i) we calculate the FM self-energy in Eq.~(\ref{eq:FM_diag_G0}) on a coarse 8$\times$8$\times$8 $\mathbf{k}$-point grid;
(ii) we add the self-energy to the bare Kohn-Sham eigenvalues on the coarse mesh;
(iii) we perform a Wannier interpolation of the bare and the corrected eigenvalues to a fine mesh;
(iv) we obtain the interpolated self-energy on the fine mesh from the difference between the bare and the corrected eigenvalues; (v) we evaluate the summation corresponding to the second term on the right hand side of
Eq.~(\ref{eq:epsilon_from_Ank}).
Following this procedure,
the total computational cost of calculating the full polaronic renormalization
of band gaps
is approximately given by 
(i) the cost of performing an adiabatic electron and hole polaron calculation
as in Ref.~\onlinecite{SioPRB2019},
plus
(ii) the cost of performing a standard AH-based band structure renormalization calculation
on a relatively coarse $\mathbf{k}$-point mesh on the Brillouin zone.

We note that since our Hamiltonian in Eq.~(\ref{eq:elph_ham}) 
only considers electron-phonon interactions to linear order in the atomic displacements,
our self-energy does not include the standard Debye-Waller (DW) contribution,\cite{AllenHeineJPC1976,GiustinoRMP2017}
and this term must be added separately to be consistent with previous work.
To evaluate this contribution, we use
the method presented in Ref.~\onlinecite{LihmPRB2020} as implemented in the {\tt PHonon} code
on a coarse Brillouin-zone mesh, and add the result to
the FM self-energy before proceeding with the interpolation procedure described above.
The dispersions of the Fan-Migdal and Debye-Waller self-energies for the valence and conduction bands 
of LiF are shown in Ref.~\onlinecite{AccompanyingPaper}.

In Figs.~\ref{fig:LiF} (c) and (d) we analyze the convergence 
of the Fan-Migdal and Debye-Waller contributions to the polaron energy with the Brillouin-zone grid
for the electron and the hole polaron, respectively.
As we discuss in Ref.~\onlinecite{AccompanyingPaper},
in the case of the large electron polaron the quasiparticle amplitudes 
are localized around the conduction band bottom,
so that relatively dense $\mathbf{k}$-meshes are needed to converge the average of the FM+DW self-energy within 1 meV.
In contrast, in the case of the small hole polaron, 
the quasiparticle amplitudes are distributed across the entire Brillouin zone,
so that coarser meshes are sufficient to achieve convergence.
The converged values for the FM+DW self-energy contribution to the polaron eigenvalue
are -0.35 eV and 0.30 eV for the electron and the hole polaron, respectively.

By combining the above contributions, we find formation energies of -0.60~eV and -2.20~eV 
for the electron and the hole polaron, respectively.
Note that the formation energy for the hole polaron is negative,
but the associated renormalization of the ionization energy and thus of the band gap is positive.
This brings the total polaronic renormalization of the band gap to \mbox{$-2.8$ eV}.
This value is considerably larger than that obtained within the Allen-Heine theory ($-1.2$~eV),
where the Fan-Migdal and Debye-Waller self-energies are evaluated at the band edges, 
without taking into account the quasiparticle amplitudes.
This result demonstrates that polaronic localization can have a significant effect on 
the band gap renormalization of solids. 
We elaborate more on this point in the companion
manuscript, Ref.~\onlinecite{AccompanyingPaper}.

\section{Localization and translational invariance}\label{sec:loc}

\def\br{{\bf r}}
\def\bR{{\bf R}}
\def\bk{{\bf k}}
\def\bq{{\bf q}}
\def\bp{{\bf p}}
\def\k{{\kappa}}
\def\a{{\alpha}}

For completeness, in this final section we address one formal question that arises in the polaron literature, 
and which pertains to the nature of the localization of a polaron in real space.\cite{Buimistrov1957a,Buimistrov1957b,Allcock1956}

The question is on how to reconcile the spatial localization of the polaron with the translational 
invariance of the Hamiltonian in Eq.~\eqref{eq:elph_ham}: Since $\hat H$ commutes with the
the lattice translation operator, the ground state must also be an eigenstate of the translation. 
This issue has already been discussed in prior literature,\cite{Allcock1956} therefore we only
touch upon those aspects that are relevant to the present work.

To clarify the relation between translational invariance and localization, we use the 
textbook example of the hydrogen atom as a proxy for an interacting electron-phonon system.
In this proxy, the proton replaces the concentration of ionic charge resulting from the formation
of the polaron.  The general expression for the eigenfunction of the hydrogen atom Hamiltonian\cite{Merzbacher1998} is:
  \begin{equation}\label{eq.h.1}
    \Psi_\bk(\br_e,\br_p) = \frac{1}{\sqrt{V}} \exp\left[i\bk\cdot \frac{m_e \br_e + m_p \br_p}{m_e+m_p}\right] \,
        \psi_{nlm}(\br_e - \br_p),
  \end{equation}
where $\br_e$ and $\br_p$ are the position of the electron and the proton, respectively, $m_e$ and $m_p$
their respective masses, $\psi_{nlm}$ is the hydrogenic eigenstate in the standard notation, 
and $V$ is the volume of the box
where the atom is contained. Since the Hamiltonian of this atom commutes with the
translation operator $\hat T_\bR$ that acts simultaneously on $\br_e$ and $\br_p$, 
$\Psi_\bk$ is also a translation eigenstate:
  \begin{equation}\label{eq.h.2}
    \hat T_\bR \Psi_\bk(\br_e,\br_p) = 
\exp(-i\bk \cdot \bR)\,\Psi_\bk(\br_e,\br_p),
  \end{equation}
as well as an eigenstate of the total momentum with eigenvalue $\hbar \bk$.
From these relations we see that the coupled electron-proton state is completely delocalized.
In particular, if we look for the probability $n(\br_e)$ of finding the electron irrespective of the location of the
proton, we have:
  \begin{equation}\label{eq.h.3}
    n(\br_e) = \int d\br_p |\Psi_\bk(\br_e,\br_p)|^2 = \frac{1}{V},
  \end{equation}
therefore the electron is fully delocalized over the box that contains the atom. On the other hand, if we consider
the conditional probability $P(\br_e| \br_p = \br_0)$ of finding the electron when the proton is located at
$\br_0$, we find:
  \begin{equation}
    P(\br_e| \br_p = \br_0) =  |\Psi_\bk(\br_e,\br_0)|^2 = \frac{1}{V} |\psi_{nlm}(\br_e - \br_0)|^2,
  \end{equation}
which is localized around $\br_0$. Similar considerations hold for excitons within the Bethe-Salpeter
formalism.\cite{OnidaRMP2002}
The situation for polarons is analogous to the above example of the hydrogen atom: electrons and atomic displacements
are localized with respect to each other, but the many-body wavefunction is delocalized in the sense
of Eq.~\eqref{eq.h.2}.

In the same way as it is convenient to study the hydrogen atom by 
using a center-of-mass reference frame,
or equivalently by
``pinning'' the center of mass
at the origin of the reference frame,
in our approach we pin the polaron at a fixed location in space. In Sec.~\ref{sec:frohlich}
this is implicitely achieved by centering the variational ansatz at $\br=0$
[cf.\ Eq.~\eqref{eq:polaron_wf_r}], and in Sec.~\ref{sec:LiF} it is achieved by initializing the polaron
wavefunction using a wavepacket at the center of the BvK supercell.

The use of polaron pinning is not mere technical expedient, it is rather
a necessity. Indeed, one limitation of the single-particle Green's function $G$ is that it
only contains electronic variables, therefore
the Dyson orbitals $f_s(\br) $ appearing in Eq.~\eqref{eq:G_Lehmann} only depend on the
electronic coordinates, unlike many-body wavefunctions such as that in Eq.~\eqref{eq.h.1}.

There are several possible avenues to overcome this limitation: (i) One could abandon the standard single-particle
Green's function 
formalism, and replace it with Green's functions for both electrons and phonons. 
This choice carries two limitations:
first, the complexity of these Green's functions grows combinatorially with the number of 
phonon modes;
second, this choice
would defeat our purpose of developing a unifying formalism that connects polaron calculations and 
many-body calculations of band structure renormalization. (ii) One could work directly with many-body wavefunctions
of electrons and phonons. This is essentially the approach taken by Pekar and coworkers in Refs.~\onlinecite{Buimistrov1957a,Buimistrov1957b},
and is amenable to incorporating translational invariance. The drawback of this approach is that it is 
a wavefunction method, hence it faces the same exponential wall that hinders direct solutions of the
many-body Schr\"odinger equation for interacting electrons. (iii) One could formally break the 
translational invariance of the Hamiltonian in Eq.~\eqref{eq:elph_ham} by introducing a small perturbation. 
Such a perturbation could be the potential of an impurity or the confining potential of a finite crystal. 

In the latter case (iii), the electron and the lattice distortion are pinned, and the symmetry-breaking perturbation can
be set to zero at the end of the calculation.
This approach is equivalent to retaining small but nonzero fictitious forces in Schwinger's functional
derivation, Eq.~\eqref{eq:source_term}.
In the present work, when we refer to polaron localization in real space, we implicitly consider that
such a small perturbation is present in the Hamiltonian as an additional term in Eq.~\eqref{eq:elph_ham},
so that translational invariance is slightly broken, localization survives, and the energetics of the polaron
is not affected.

\section{Summary and outlook} \label{sec:conclusion}

In summary, we have presented an \textit{ab initio} Green's function theory of polarons,
which unifies the perturbative weak-coupling approach and the adiabatic strong-coupling approach
to the polaron problem.
Starting from a general electron-phonon Hamiltonian,
we have derived a Dyson equation for the electron Green's function,
accounting for possible static displacements of the atomic nuclei in the ground state of the system with
an excess electron or hole.
In addition to the conventional Fan-Midgal dynamical self-energy, we identified a new
self-energy contribution which results from static lattice distortions in the polaron state.
After presenting the general formalism, 
we have outlined several approximations that enable practical implementations of the theory 
in current \textit{ab initio} software.
This analysis establishes unambiguously the links between our formalism,
density functional calculations of polarons,\cite{SioPRL2019,SioPRB2019}
and the Allen-Heine theory of band structure renormalization.\cite{AllenHeineJPC1976}

In order to benchmark our method,
we have studied the ground state energy of the Fr{\"o}hlich polaron, and found that our
approach is in very good agreement with Feynman's path-integral solution and with diagrammatic
Monte Carlo calculations, at all coupling strengths.
As a first \textit{ab initio} calculation using this method,
we have computed the polaronic band gap renormalization in LiF.
The main results and implications of our \textit{ab initio} calculations are discussed in the companion
manuscript.\cite{AccompanyingPaper}

The agreement between our theory and previous diagrammatic Monte Carlo calculations for the
Fr\"ohlich model might appear suprising. In fact these previous calculations involve summations
over a very large number of electron-phonon self-energy diagrams,\cite{ProkofevPRL1998,MischenkoPRB2000,HahnPRB2018}
while only two self-energies are considered in this work. 
The main difference between our approach and the diagrammatic Monte Carlo method is that 
in our case the sum over all electron-phonon diagrams is replaced by a set of self-consistent equations 
defining the exact interacting Green's function.
This strategy allows us to describe localization effects,
which become dominant at strong coupling, via the self-consistent polaronic self-energy given in Eq.~(\ref{eq:LP_from_G}).
The remaining non-adiabatic electron-phonon interactions are encoded in the FM self-energy given in
Eq.~(\ref{eq:selfen_FM}),
whose lowest-order approximation is enough to capture the renormalization of large polarons at weak coupling.
Higher-order diagrams could be included via the vertex function $\Gamma$, 
but on the basis of the results presented in this work we expect their contribution to be small.
It is possible that the inclusion of vertex corrections will further reduce
the slight deviation between our present results and diagrammatic Monte Carlo calculations.

The success of our self-consistent many-body approach is reminiscent of Hedin's GW equations for
the electron-electron problem.\cite{HedinPR1965,HedinLundqvistSSP1969}
By converting the infinite sum of diagrams for the bare Coulomb interaction into a set of self-consistent equations,
it was found
that the electron-electron self-energy could be expanded in terms of the screened Coulomb interaction,
and this strategy proved highly successful over the past four decades.\cite{HybertsenLouiePRB1986,
OnidaRMP2002,ReiningWIREs2018,GolzeFC2019} 
In the same spirit, in the present work we employed the
functional derivative technique of Schwinger to replace a summation over infinite electron-phonon diagrams
into the self-consistent solution of a set of equation for the electron Green's function and the interaction
self-energies. This strategy allowed us to show that
adiabatic localization and dynamical many-body effects
are not separate and inconsistent ways to look at the electron-phonon problem. 
Rather,
both contributions are complementary aspect of the same problem, and need to be taken into accound on the
same footing.

Many improvements upon the present method are possible. For example, in this work we mostly focused
on perturbative solutions of the self-consistent many-body polaron equations; in the future it will be
interesting to test full-blown self-consistent schemes for better accuracy. Furthermore, in this work
we only focus on the polaron ground state, but the formalism contains information about excited states
as well; work on polaron excitations would 
be useful to investigate finite-temperature properties
and the response of polarons to external fields. 
Another interesting development would be to calculate the renormalization 
of the phonon Green's function on the same footing as the electron Green's 
function,
which would require 
upgrading the starting point in Eq.~(\ref{eq:elph_ham})
to a more general electron-ion Hamiltonian.
\cite{BaymAP1961,GiustinoRMP2017}
This further step would allow us to investigate the signature of polarons in vibrational
spectroscopy via the change in the phonon frequencies.\cite{MiyakeJPSJ1976}
Lastly, systematic calculations using the present approach for a broad library of materials will
be needed to assess the significance of polaronic effects, and their role in the phonon-induced renormalization
of the band structure of solids.

We hope that this work will be useful as a starting point to investigate polarons in real materials
from the point of view of \textit{ab initio} many-body methods.

\begin{acknowledgments}
This research is primarily supported by the Computational Materials Sciences Program funded by the U.S. Department of 
Energy, Office of Science, Basic Energy Sciences, under Award No. DE-SC0020129 (JLB, CL, WHS: formalism, software 
development, \textit{ab initio} calculations, manuscript preparation), and by the National Science Foundation, 
Office of Advanced Cyberinfrastructure under Grant No. 2103991 (FG: project conception and supervision, manuscript 
preparation). The authors acknowledge the Texas Advanced Computing Center (TACC) at The University of Texas 
at Austin for providing HPC resources, including the Frontera and Lonestar5 systems, that have contributed 
to the research results reported within this paper.  URL: http://www.tacc.utexas.edu.  This research used 
resources of the National Energy Research Scientific Computing Center, a DOE Office of Science User Facility 
supported by the Office of Science of the U.S.  Department of Energy under Contract No. DE-AC02-05CH11231. 
WHS was supported by the Science and Technology Development Fund of Macau SAR (under Grants No. 0102/2019/A2) and the LvLiang Cloud Computing Center of China for providing extra HPC resources, including the TianHe-2 systems.
IGG and AE acknowledge the Department of Education, Universities and Research of the Eusko Jaurlaritza and the University of the Basque Country UPV/EHU (Grant No. IT1260-19), 
the Spanish Ministry of Economy and Competitiveness MINECO (Grants No. FIS2016-75862-P and No.
PID2019-103910GB-I00),
and the University of the Basque Country UPV/EHU (Grant No. GIU18/138) for financial support.
JLB acknowledges UPV/EHU (Grant No. PIF/UPV/16/240), MINECO (Grant No. FIS2016-75862-P) and DIPC for financial support in the initial stages of this work.
\end{acknowledgments}

\bibliography{bibliography}

\end{document}